\documentclass[%
 reprint,
superscriptaddress,
 amsmath,amssymb,
pre,
]{revtex4-2}
\usepackage[
  compatibility=false,
  labelfont=bf
]{caption}

\usepackage{subcaption}
\usepackage{ragged2e}

\usepackage[percent]{overpic}
\usepackage{xcolor}
\usepackage{graphicx}% Include figure files
\usepackage{dcolumn}% Align table columns on decimal point
\usepackage{bm}% bold math
\usepackage[normalem]{ulem} %THIS PACKAGE MESSES UP THE BIBLIOGRAPHY, DO NOT UNCOMMENT

\usepackage{hyperref}% add hypertext capabilities
\begin{document}

%\preprint{APS/123-QED}

\title{Predicting Plasticity in Two-Dimensional Foam Channel Flow Around an Obstacle}

\author{Alexandre Stepanetz}
\affiliation{Univ. Grenoble Alpes, CNRS, LIPhy, 38000 Grenoble, France}

\author{Bahaa Mazloum}
\affiliation{Univ. Grenoble Alpes, CNRS, LIPhy, 38000 Grenoble, France}

\author{Benjamin Dollet}
\affiliation{Univ. Grenoble Alpes, CNRS, LIPhy, 38000 Grenoble, France}

\author{Misaki Ozawa}
\affiliation{Univ. Grenoble Alpes, CNRS, LIPhy, 38000 Grenoble, France}

\date{\today}% It is always \today, today,

\begin{abstract}

We study the prediction of plastic activity in the confined channel flow of two-dimensional amorphous soft particles around a circular obstacle. Using datasets generated with a particle-based bubble model, we formulate the prediction problem within two supervised-learning frameworks: regression of the non-affine displacement and binary classification of neighbor change events. A key technical challenge is that the obstacle and the confining walls explicitly break translational and rotational symmetries. We address this issue by introducing additional structural descriptors that encode the positions of particles relative to these boundaries. Starting from simple linear models, we systematically increase the complexity of the learning framework by considering a logarithmic transformation of the target variable, the incorporation of particle-size information, the addition of symmetry-breaking obstacle and wall descriptors, the coarse-graining of local structural descriptors, and nonlinear neural-network models. We find that the obstacle and wall descriptors provide the largest improvement in predictive performance. Nevertheless, the models considered here capture mainly the overall localization of plastic activity near the obstacle and do not fully reproduce its detailed heterogeneous pattern in individual configurations. A perturbation analysis indicates that this heterogeneity is robustly encoded in the initial structure, suggesting that further progress requires more expressive structural representations and machine-learning architectures.

\end{abstract}

\maketitle

%\tableofcontents

\section{Introduction}

Foam dynamics plays an important role in soft matter physics, not only because of its practical relevance in applications, but also because it provides a useful model system to study how constituent particles undergo plastic rearrangements. 
Indeed, early work by Bragg \& Nye~\cite{Bragg1947} and Argon \& Kuo~\cite{Argon1979} used bubble rafts as an analogue system to gain microscopic insight into particle-scale plastic rearrangements in metallic glasses.

Plastic deformation in foams is typically induced by external loading, such as simple shear, compression, or confined channel flow. 
In particular, confined channel flow around a circular obstacle provides a fundamental geometry to investigate bubble dynamics~\cite{dollet2007two}, in analogy with the classical Stokes-resistance problem in continuum hydrodynamics. 
Recently, we studied this geometry using molecular dynamics simulations of a two-dimensional bubble model, systematically varying control parameters such as the packing fraction, the polydispersity, and the magnitude of the external driving force~\cite{mazloum2026channel}. 
This study revealed several interesting phenomena, including a crossover from crystalline-like dislocation gliding motion~\cite{sethna2017deformation} to amorphous-like localized rearrangements~\cite{bonn2017yield,nicolas2018deformation} with increasing polydispersity, as well as a yield-drag transition~\cite{Cantat2006,Raufaste2007} controlled by the external driving force and packing fraction. 
In particular, plastic activity in this system occurs in a highly heterogeneous and intermittent manner. 
For crystalline or weakly polydisperse systems, defects such as dislocations can be visually identified, and the subsequent plastic motion often follows well-defined sliding directions~\cite{sethna2017deformation,ghimenti2024shear}. 
In this sense, the relation between structure and dynamics is relatively visible. 
In contrast, in amorphous systems, the static structure appears disordered and featureless, at least to the naked eye, while the subsequent dynamics displays strong spatial heterogeneity. 
This behavior is a hallmark of amorphous and glassy materials, commonly referred to as dynamical heterogeneity~\cite{berthier2011dynamical,karmakar2014growing}. 
Predicting the future plastic activity that gives rise to such dynamical heterogeneity from a static amorphous configuration is therefore a challenging task~\cite{manning,cubuk2015identifying,richard2020predicting}. 
It is closely related to the long-standing problem of identifying structural order parameters in amorphous materials that are relevant for plasticity and flow~\cite{coslovich2007understanding,royall2015role,tanaka2019revealing}. 

Recent advances in machine-learning techniques have had a major impact on the prediction of dynamics from static snapshots. 
In particular, the glass-physics community has made substantial progress in predicting the future mobility of particles in glass-forming liquids using supervised learning~\cite{jung2025roadmap}. 
In this approach, a static configuration, or structural descriptors computed from it, is used as the input feature, while a dynamical quantity such as particle mobility is used as the target variable. 
A wide range of machine-learning methods has been applied to this problem, from simple linear regression and support vector machines to state-of-the-art graph neural networks~\cite{cubuk2015identifying,bapst2020unveiling,boattini2021averaging,shiba2023botan,pezzicoli2024rotation}. 
These studies have shown that machine learning can predict the future dynamics of glass-forming liquids with high accuracy.

However, as the complexity of the machine-learning model increases, for example in graph neural networks, it becomes more difficult to extract physical insight from the model. 
In other words, accurate prediction does not necessarily imply physical understanding~\cite{teney2022predicting,swain2024machine,sharma2026interpretability}. 
Remarkably, Filion and coworkers revisited the success of graph neural networks and argued that their predictive performance can largely be understood as arising from an effective coarse-graining of local structural information~\cite{boattini2021averaging,alkemade2022comparing}. Building on this insight, they showed that a much simpler machine-learning approach, namely, linear regression using coarse-grained, hand-crafted structural descriptors, can achieve predictive accuracy comparable to that of graph neural networks~\cite{boattini2021averaging,alkemade2022comparing}.
This is an important result because it combines high predictive performance with a level of interpretability that is difficult to obtain in deep-learning approaches.

In this paper, motivated by this strategy, we study the machine-learning prediction of amorphous foam dynamics using relatively simple and interpretable models. 
Rather than starting directly from a highly complex architecture, we incrementally increase the complexity of the model and monitor how the prediction performance improves. 
This strategy allows us to identify which structural ingredients are essential for predicting future plastic activity. 
We believe that such an approach can provide not only accurate predictions, but also physical insight into the structural origin of heterogeneous dynamics in amorphous foams.

One of the key technical challenges in this study is how to account for the breaking of translational and rotational symmetries caused by the presence of the obstacle and the confining walls in the channel-flow geometry. In previous machine-learning studies of glass-forming liquids and driven amorphous materials, systems with translational and rotational symmetries were typically considered, and these symmetries were exploited in the design of structural descriptors and neural-network architectures~\cite{richard2020predicting,jung2025roadmap}. In general, incorporating symmetries into machine-learning architectures reduces the number of parameters and improves generalization.

However, in channel foam flow around an obstacle, plasticity is induced by interactions with the obstacle, leading to heterogeneous dynamics in its vicinity, as demonstrated experimentally~\cite{dollet2007two} and numerically~\cite{mazloum2026channel}. This leads to an explicit breaking of translational and rotational symmetries, in stark contrast to previous studies of glass-forming liquids and driven amorphous materials. We will address this technical problem by introducing additional features that explicitly encode the symmetry breaking, on top of conventional descriptors that are translationally and rotationally invariant~\cite{behler2007generalized}. This aspect constitutes a new approach from a technical point of view.

The paper is organized as follows. Section~\ref{sec:simulation_dataset} describes the simulation model and the construction of the dataset. Section~\ref{sec:ML} presents the prediction of plasticity using regression and classification approaches. Section~\ref{sec:perturbation} discusses how much information about the future dynamics is encoded in the initial configuration. Finally, we conclude and discuss future perspectives in Sec.~\ref{sec:conclusion}.

\section{Simulation model and dataset}
\label{sec:simulation_dataset}

We perform molecular simulations using the model developed in Ref.~\cite{mazloum2026channel}, where
two-dimensional foam flow around a circular obstacle was studied using a
particle-based bubble model. This setup was studied experimentally in Ref.~\cite{dollet2007two}. Here, we briefly summarize the simulation model
and describe the construction of the dataset used for the machine-learning
analysis in the next section.

\subsection{Simulation model}

\begin{figure*}[htbp]
    \centering
    \begin{subfigure}[t]{0.32\linewidth}
        \captionsetup{justification=raggedright, singlelinecheck=false, position=above}  
        \caption{$D_{\rm min}^2$ for $\Delta t=100$}
        \includegraphics[width=\linewidth]{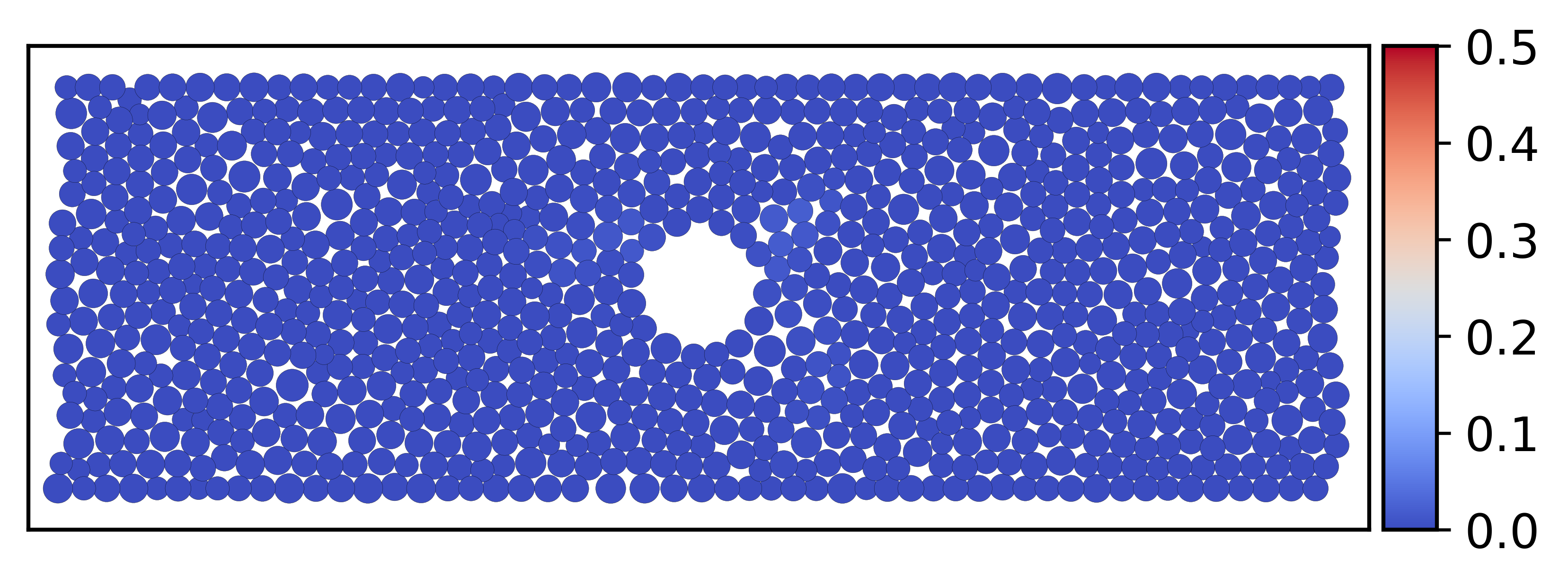}
        \label{fig:delta10a}
    \end{subfigure}
    \hfill
    \begin{subfigure}[t]{0.32\linewidth}
        \captionsetup{justification=Justified, singlelinecheck=false, position=above}
        \caption{$D_{\rm min}^2$ for $\Delta t=600$}
        \includegraphics[width=\linewidth]{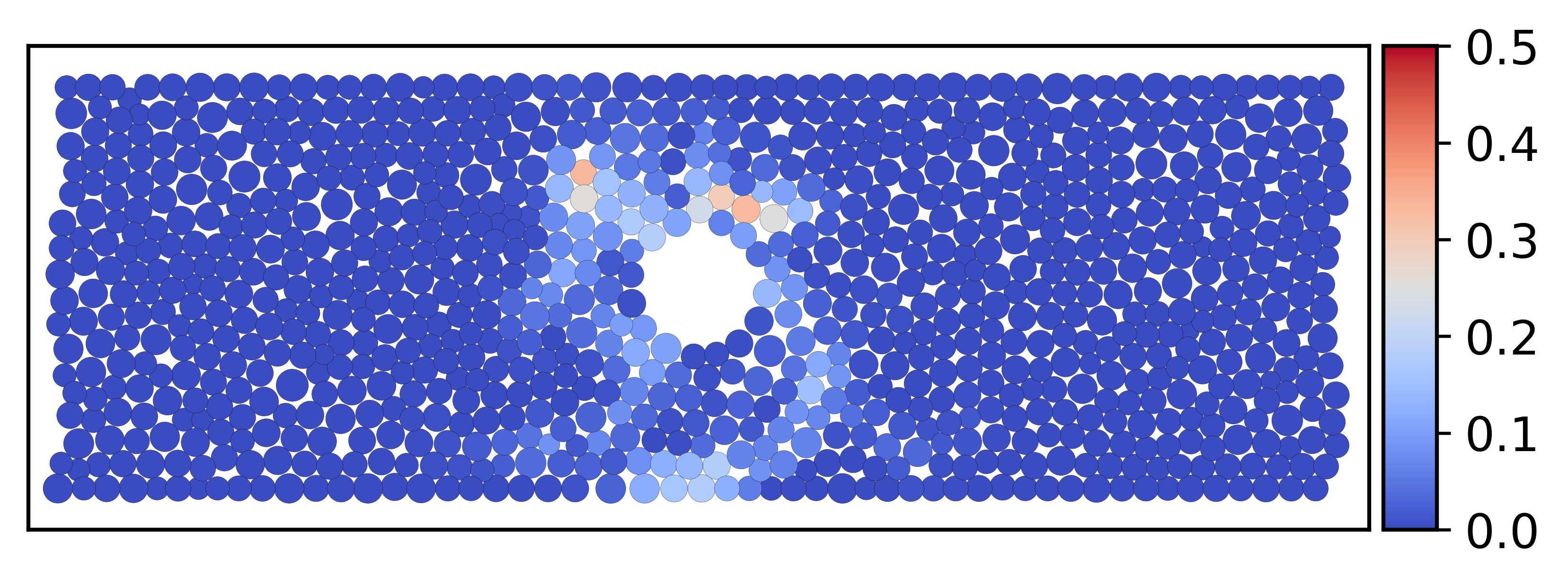}
        \label{fig:delta10b}
    \end{subfigure}
    \hfill
    \begin{subfigure}[t]{0.32\linewidth}
        \captionsetup{justification=Justified, singlelinecheck=false, position=above}
        \caption{$D_{\rm min}^2$ for $\Delta t=1200$}
        \includegraphics[width=\linewidth]{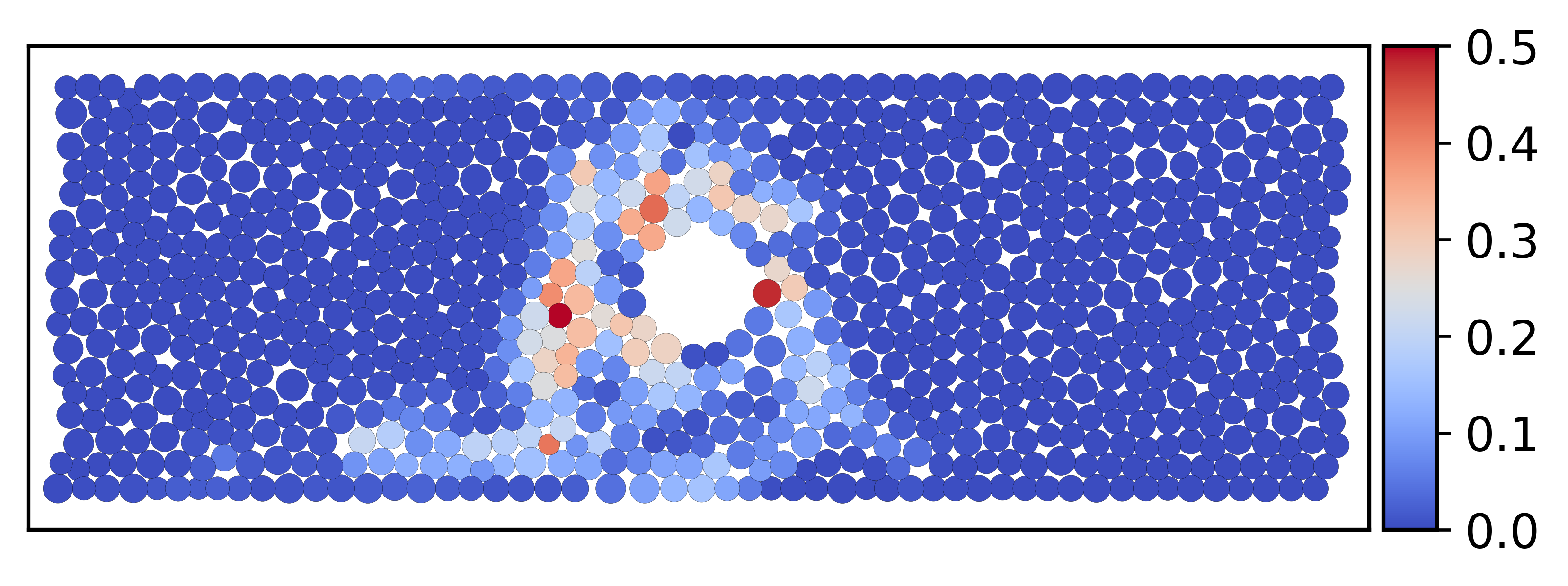}
        \label{fig:delta10c}
    \end{subfigure}
    \vspace{3mm}
    \begin{subfigure}[t]{0.32\linewidth}
        \captionsetup{justification=Justified, singlelinecheck=false, position=above}
        \caption{$\log(D_{\rm min}^2)$ for $\Delta t=100$}
        \includegraphics[width=\linewidth]{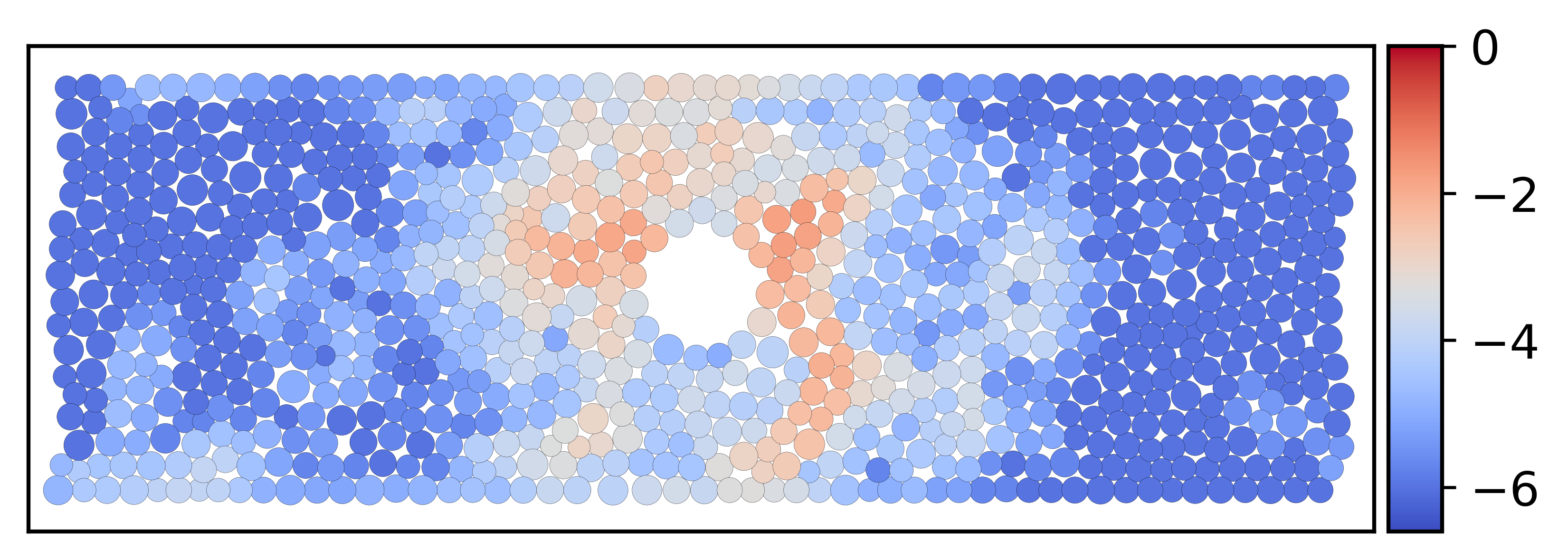}
        \label{fig:delta30a}
    \end{subfigure}
    \hfill
    \begin{subfigure}[t]{0.32\linewidth}
        \captionsetup{justification=Justified, singlelinecheck=false, position=above}
        \caption{$\log(D_{\rm min}^2)$ for $\Delta t=600$}
        \includegraphics[width=\linewidth]{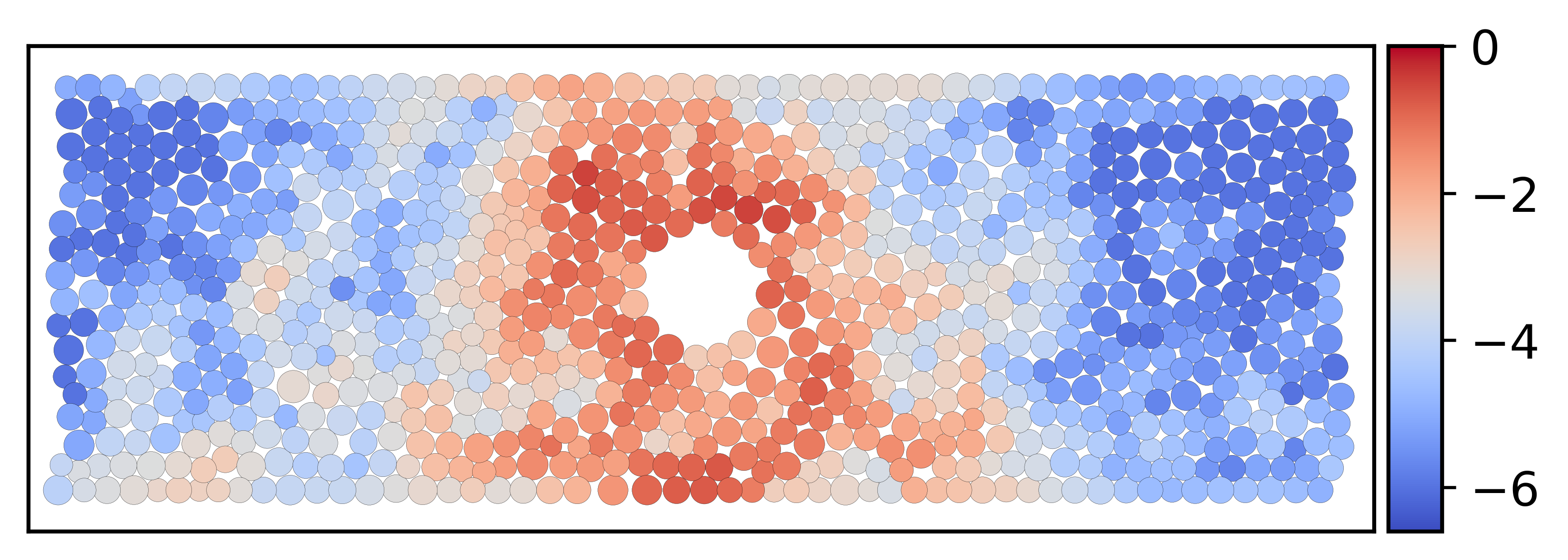}
        \label{fig:delta30b}
    \end{subfigure}
    \hfill
    \begin{subfigure}[t]{0.32\linewidth}
        \captionsetup{justification=Justified, singlelinecheck=false, position=above}
        \caption{$\log(D_{\rm min}^2)$ for $\Delta t=1200$}
        \includegraphics[width=\linewidth]{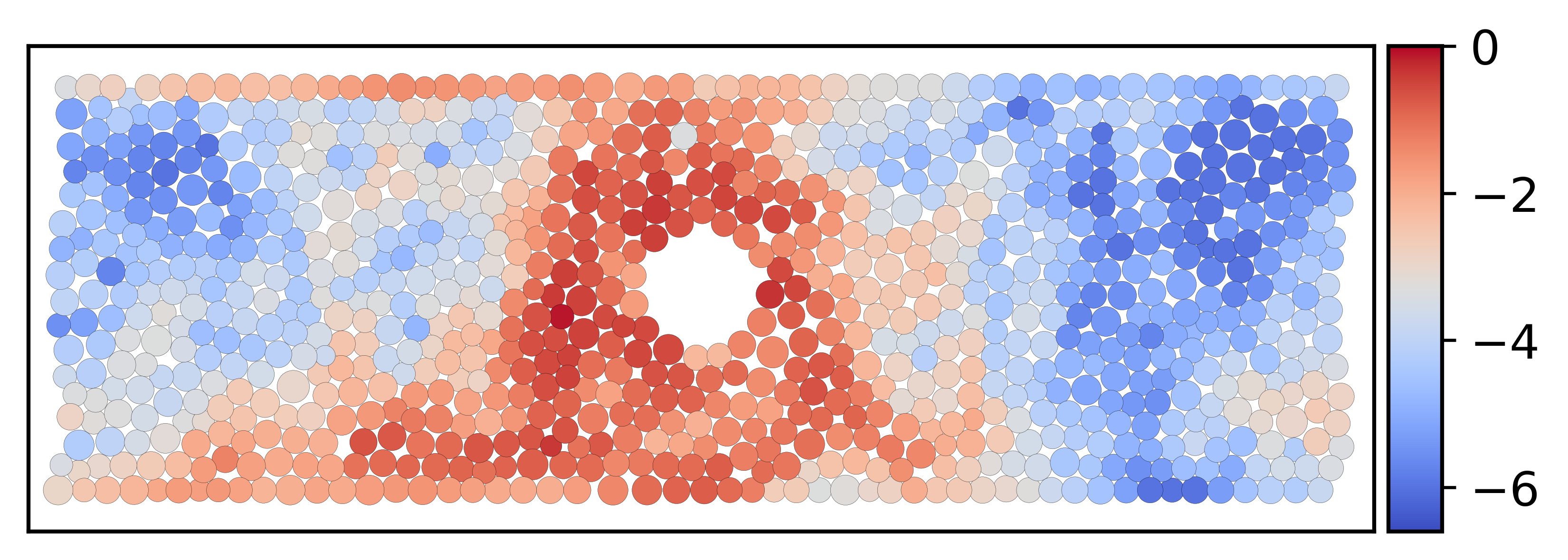}
        \label{fig:delta30c}
    \end{subfigure}
    \vspace{3mm}
    \begin{subfigure}[t]{0.32\linewidth}
        \captionsetup{justification=Justified, singlelinecheck=false, position=above}
        \caption{Neighbor change events for $\Delta t=100$}
        \includegraphics[width=\linewidth]{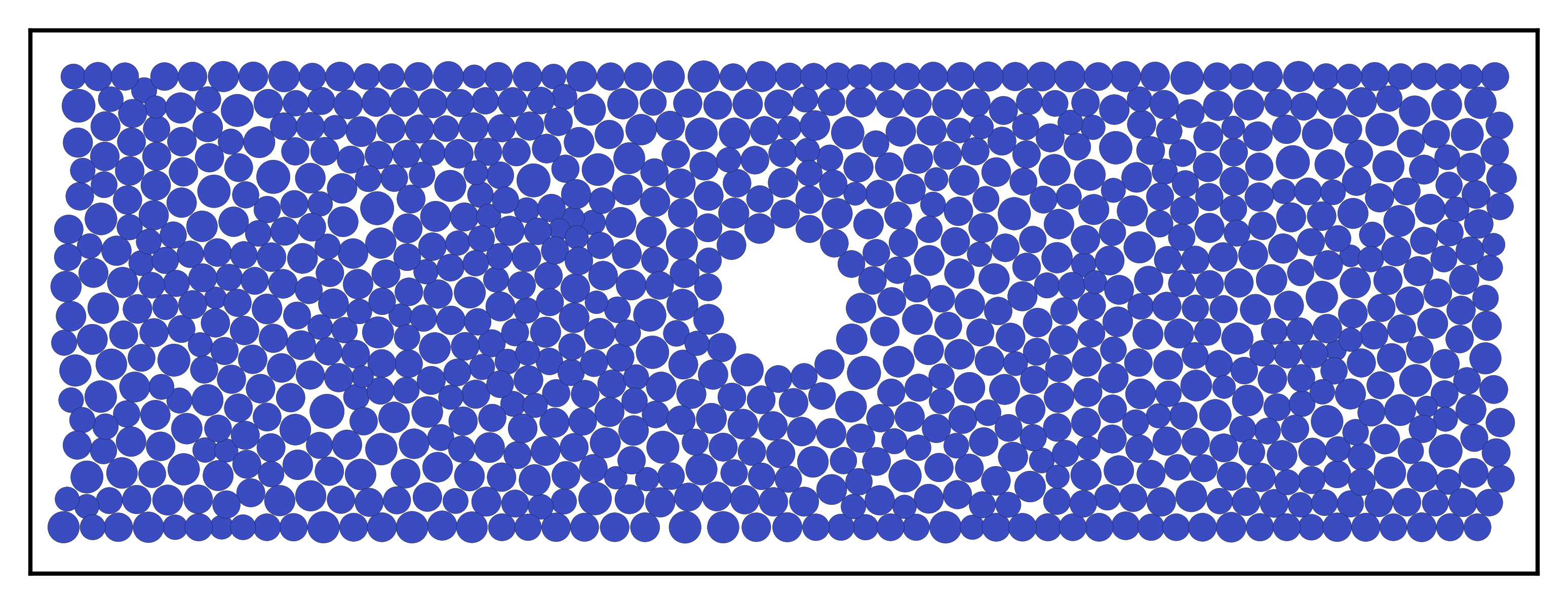}
        \label{fig:delta120a}
    \end{subfigure}
    \hfill
    \begin{subfigure}[t]{0.32\linewidth}
        \captionsetup{justification=Justified, singlelinecheck=false, position=above}
        \caption{Neighbor change events for $\Delta t=600$}
        \includegraphics[width=\linewidth]{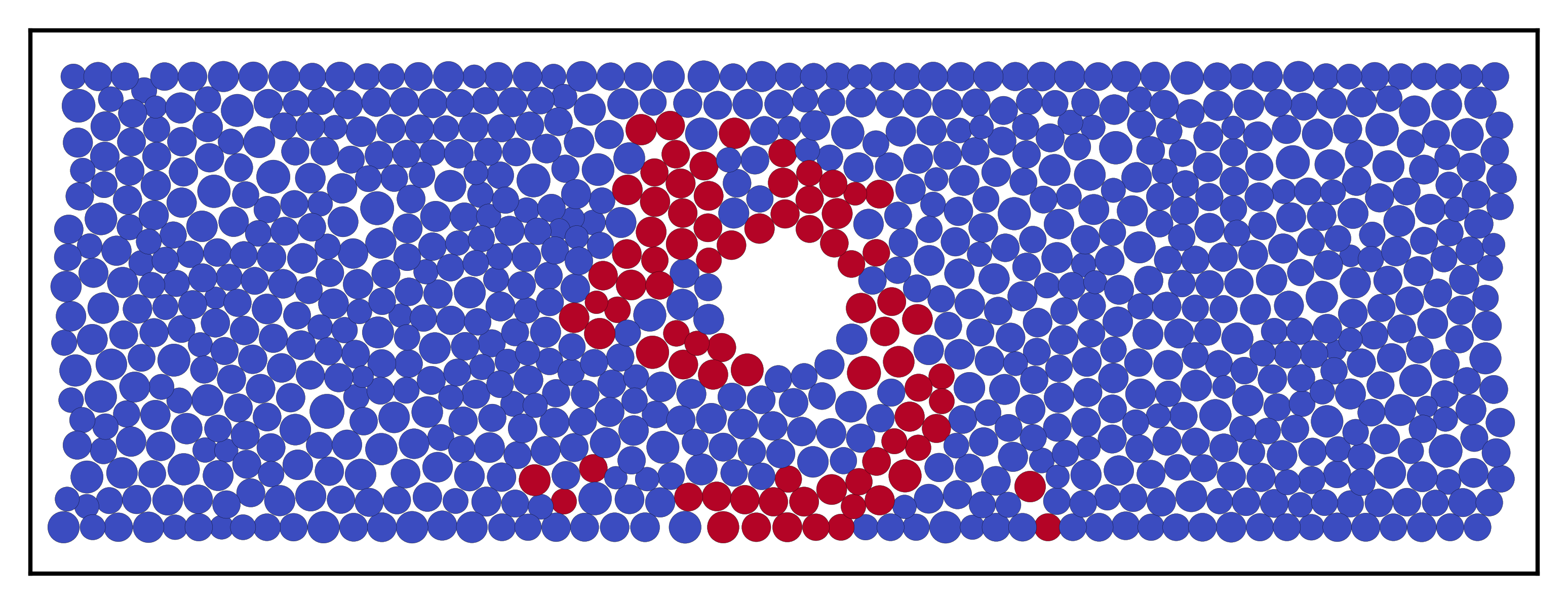}
        \label{fig:delta120b}
    \end{subfigure}
    \hfill
    \begin{subfigure}[t]{0.32\linewidth}
        \captionsetup{justification=Justified, singlelinecheck=false, position=above}
        \caption{Neighbor change events for $\Delta t=1200$}
        \includegraphics[width=\linewidth]{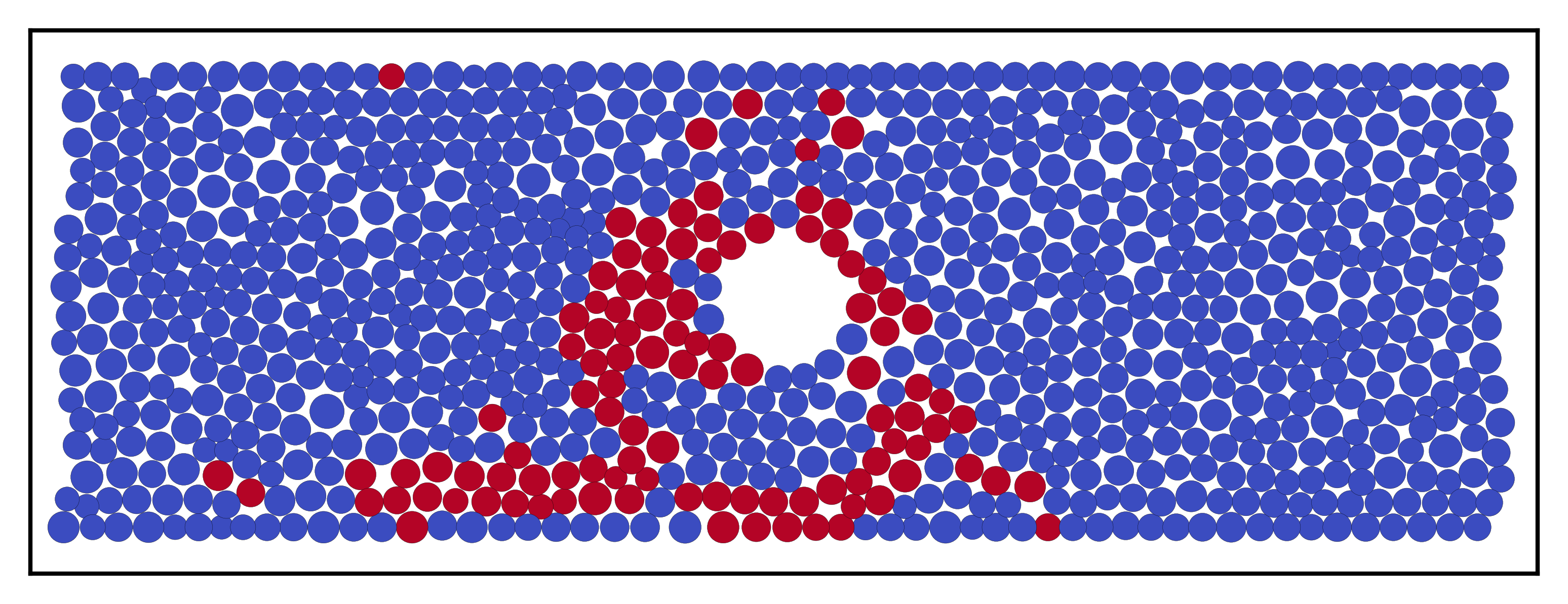}
        \label{fig:delta120c}
    \end{subfigure}
    \caption{(a--c) Time evolution of the non-affine displacement $D_{\rm min}^2$ for
$\Delta t=100$ (a), $\Delta t=600$ (b), and $\Delta t=1200$ (c).
(d--f) Corresponding plots of $\log D_{\rm min}^2$ for
$\Delta t=100$ (d), $\Delta t=600$ (e), and $\Delta t=1200$ (f).
(g--i) Corresponding plots of neighbor change events. Red and blue particles
indicate particles that undergo and do not undergo a neighbor change event,
respectively.}
    \label{fig:dataset}
\end{figure*}

The system consists of $N=900$ soft disks confined in a two-dimensional
rectangular channel of size $L_x \times L_y$. The top and bottom boundaries are
modeled as repulsive walls of thickness $w=0.5$, while periodic boundary
conditions are imposed along the flow direction $x$. A circular obstacle of
diameter $\sigma_{\rm obs}=10$ is placed at the center of the channel. The box
aspect ratio is fixed as $L_x=\gamma L_y$ with $\gamma=3$, and $L_y$ is adjusted
to control the packing fraction.

The disks are polydisperse. The degree of polydispersity is quantified by $\delta =
\sqrt{\overline{\sigma^2}-\overline{\sigma}^2}/\overline{\sigma}$
where $\sigma_i$ is the diameter of particle $i$ and
$\overline{\sigma}=N^{-1}\sum_i \sigma_i$ is the mean diameter.
We set the polydispersity to $\delta=0.15$ throughout this paper, for which
the system forms amorphous configurations.
The packing fraction $\phi$ is defined using the area accessible to the
particles, and we set $\phi=1.2$ throughout this paper.

The particle dynamics is overdamped~\cite{durian1995foam}. The position
${\bf r}_i(t)=(x_i(t),y_i(t))$ of particle $i$ evolves as
\begin{equation}
    \zeta \frac{d{\bf r}_i}{dt}
    =
    {\bf f}^{\rm int}_i
    +
    {\bf f}^{\rm wall}_i
    +
    {\bf f}^{\rm obs}_i
    +
    {\bf f}^{\rm ext},
    \label{eq:overdamped_foam}
\end{equation}
where ${\bf f}^{\rm int}_i$ is the interaction force from other particles,
${\bf f}^{\rm wall}_i$ is the force from the confining walls,
${\bf f}^{\rm obs}_i$ is the force from the central obstacle, and
${\bf f}^{\rm ext}$ is a uniform external driving force. We set
$\zeta=1$. The driving force is applied along the channel direction, ${\bf f}^{\rm ext}=(f^{\rm ext},0)$.
We set $f^{\rm ext}=0.001$ in this study, corresponding to a slow-driving regime in
which plastic rearrangements can be identified clearly.

Particles interact through a purely repulsive finite-range potential,
\begin{equation}
    v^{\rm int}_{ij}(r_{ij})
    =
    \frac{\epsilon}{\alpha}
    \left(1-\frac{r_{ij}}{\sigma_{ij}}\right)^{\alpha}
    \Theta(\sigma_{ij}-r_{ij}),
    \label{eq:pair_potential}
\end{equation}
where $r_{ij}=|{\bf r}_i-{\bf r}_j|$,
$\sigma_{ij}=(\sigma_i+\sigma_j)/2$, and $\Theta$ is the Heaviside step
function. In this work, we use the Hertzian exponent $\alpha=5/2$. The walls and
the obstacle are also modeled by short-range harmonic repulsions with stiffness
$K=10$. Length, time, and energy are measured in units of
$\overline{\sigma}$, $t_0=\zeta \overline{\sigma}^{\,2}/\epsilon$, and
$\epsilon$, respectively. Equation~\eqref{eq:overdamped_foam} is integrated
using the Euler method with time step $dt=0.1$.

The simulations are initialized from random configurations. After an initial
transient, the system reaches a statistically steady state, where the potential
energy fluctuates around a stationary value. The machine-learning dataset is
constructed only from this steady-state regime.

\subsection{Dataset construction}

The aim of the machine-learning task is to predict future plastic activity from
a static snapshot~\cite{cubuk2015identifying,richard2020predicting}. Each data point corresponds to one particle in one
configuration. The input features are structural descriptors (see Sec.~\ref{sec:ML}) computed from the
instantaneous configuration at time $t$, while the target variable is computed
from the subsequent dynamics between $t$ and $t+\Delta t$. 

We consider two related prediction tasks. The first one is a regression task,
where the target is the local non-affine displacement measured by
$D^2_{\rm min}$~\cite{falk1998dynamics}. For particle $i$, this quantity is defined as
\begin{eqnarray}
    D^2_{{\rm min},i}(t,\Delta t)
    &=&
    \frac{1}{n_i}
    \sum_{j\in\mathcal{N}_i}
    \left|
    \left[
    {\bf r}_j(t+\Delta t)-{\bf r}_i(t+\Delta t)
    \right]
    \right.
    \nonumber \\
    &&
    \left.
    -
    (I+E^*)
    \left[
    {\bf r}_j(t)-{\bf r}_i(t)
    \right]
    \right|^2 .
    \label{eq:D2min}
\end{eqnarray}
Here, $\mathcal{N}_i$ is the set of neighbors of particle $i$ at time $t$,
$n_i$ is the number of such neighbors, $I$ is the identity matrix, and $E^*$ is
the best-fit local affine deformation tensor. Thus, $D^2_{\rm min}$ measures the
part of the particle motion that cannot be described by a local affine deformation. We define the neighbors of particle $i$ as the particles located within a cutoff distance $r_{\rm min}$ at time $t$, where $r_{\rm min}$ is chosen as the position of the first minimum of the radial distribution function.
Figure~\ref{fig:dataset}(a--c) shows the time evolution of the spatial map of
$D^2_{\rm min}$ starting from a given configuration. Plastic activity first
appears near the obstacle and then spreads with increasing $\Delta t$ in a
spatially heterogeneous manner. In Fig.~\ref{fig:dataset}(d--f), we show the
same data on a logarithmic scale, $\log(D^2_{\rm min})$, which makes the
mobile regions easier to visualize. Throughout this paper, logarithms are taken with base 10 unless otherwise
specified.

The second task is a binary classification task, where the target indicates
whether particle $i$ undergoes a plastic rearrangement between $t$ and
$t+\Delta t$. We use a neighbor change indicator,
\begin{equation}
    I_i
    =
    \begin{cases}
    1, & \text{if particle } i \text{ changes its neighbor list},\\
    0, & \text{otherwise}.
    \end{cases}
    \label{eq:T1_indicator}
\end{equation}
This definition includes the usual fourfold neighbor swapping (T1) event as a special case, but is more
general and more robust for the present soft-disk simulations. Neighbor lists
are defined using the first minimum $r_{\rm min}$ of the radial distribution
function. To reduce spurious events caused by small fluctuations, we use two
slightly different cutoffs for bond breaking and bond formation, separated by a
small margin $\delta r=0.2$~\cite{nishikawa2022relaxation,takaha2025avalanche}.
In Fig.~\ref{fig:dataset}(g--i), we show the time evolution of the
neighbor change event indicator. The events first appear near the obstacle and
then propagate as time increases.

We repeat this dataset construction for several prediction timescales
$\Delta t=50, 100, 200, 300, 600, 900, 1200$, which allows us to test how far into the future plastic activity
can be predicted from a static structure. 
We prepare $N_{\rm conf}=1000$ configurations for each value of $\Delta t$.
Thus, we have $N N_{\rm conf}=900000$ data points, each corresponding to one particle. 
These data points are split into training, validation, and test datasets.

\section{Machine learning prediction}
\label{sec:ML}

Based on the dataset presented in Sec.~\ref{sec:simulation_dataset}, we perform machine-learning prediction using both regression and classification approaches. 
All machine-learning experiments are performed using scikit-learn~\cite{pedregosa2011scikit}.
%, and the data analysis is carried out in Python. The source code and dataset are publicly available, as described in the Data Availability statement.

\subsection{Input features}

We first compute input features from the particle configurations using Behler--Parrinello structural descriptors~\cite{behler2007generalized}, which have been used to predict plastic rearrangements in sheared amorphous solids~\cite{cubuk2015identifying,rocks2021learning}. These descriptors transform particle coordinates into a set of translationally and rotationally invariant functions that characterize the local structural environment around each particle. By varying the parameters of these functions, the descriptors are designed to probe structural information over a broad range of length scales and local geometries.

The Behler--Parrinello descriptors consist of radial and angular components. The radial descriptors characterize the distribution of neighboring particles as a function of distance, whereas the angular descriptors encode orientational information associated with triplets of particles. The detailed functional forms are given in Appendix~\ref{sec:BP}. As a first step, we ignore particle-size information and treat all particles in the polydisperse system as belonging to a single species.

Thus, in total, we obtain $M=70$ features for each particle. 
Each feature is normalized to have zero mean and unit variance. 
After this normalization, the features define the feature vector of particle $i$ as
\begin{equation}
    {\bf X}_i =
    \left(1, 
    X_i^{(1)}, X_i^{(2)}, \ldots, X_i^{(M)}
    \right)^T,
\end{equation}
where $T$ is the transpose.
The first component, which is set to unity, is introduced to represent the intercept term in the linear model, as described below.
This feature vector is used as the input for the regression and classification tasks described below.

\subsection{$D_{\rm min}^2$ prediction}

We first consider the prediction of plastic activity characterized by $D_{\rm min}^2$ using a linear regression model. In this setup, the dataset consists of input feature vectors ${\bf X}_i$ and target variables $Y_i$, where $Y_i$ corresponds to the value of $D_{\rm min}^2$ for particle $i$. 

We consider the following linear model for the prediction $\hat Y_i$:
\begin{equation}
    \hat Y_i = {\bf w}^{\rm T} {\bf X}_i ,
    \label{eq:linear_model}
\end{equation}
where
\[
{\bf w} =
\left(w^{(0)},
w^{(1)}, w^{(2)}, \ldots, w^{(M)}
\right)^T
\]
is the weight vector.
Here, $w^{(0)}$ corresponds to the intercept term.
The weight vector is determined by minimizing the loss function
\begin{equation}
    \mathcal{L}({\bf w})
    =
    \frac{1}{2}
    \sum_{i=1}^{N_{\rm train}}
    \left|
    \hat Y_i - Y_i
    \right|^2
    +
    \frac{\alpha}{2} \sum_{f=1}^M
    \left(w^{(f)}\right)^2 ,
    \label{eq:loss_MSE}
\end{equation}
where the second term is the Ridge regularization term with regularization strength $\alpha$. Here, $N_{\rm train}$ denotes the number of data points in the training dataset.

To evaluate the prediction performance, we use the Pearson correlation coefficient $\rho_{A,B}$, which quantifies the linear correlation between two variables $A$ and $B$:
\begin{equation}
    \rho_{A,B}
    =
    \frac{1}{n}
    \sum_{i=1}^{n}
    \frac{
    (A_i-\mu_A)(B_i-\mu_B)
    }{
    \sigma_A \sigma_B
    } ,
\end{equation}
where $\mu_A$ and $\mu_B$ are the means of $A_i$ and $B_i$, respectively, $\sigma_A$ and $\sigma_B$ are their standard deviations, and $n$ is the number of data points.
In particular, we compute the Pearson correlation coefficient between the predicted and true values of the target variable, $\rho_{\hat Y,Y}$.
A value of $\rho_{\hat Y,Y}\approx 1$ indicates an excellent prediction, whereas $\rho_{\hat Y,Y}\approx 0$ indicates poor predictive performance.

The training, validation, and test split is performed at the configuration level. 
We use $K$-fold cross-validation to select the model hyperparameters. Specifically, the data are divided into $K$ subsets; each subset is used once for validation, while the remaining $K-1$ subsets are used for training.
We confirmed that the regularization parameter $\alpha$ has little effect on the prediction performance.
We therefore set $\alpha=0$ throughout this analysis.
To assess possible overfitting, we also computed training curves by varying the number of training configurations, as shown in Appendix~\ref{sec:training_curves}.
The close agreement between the training and validation performances confirms that the dataset contains enough data points to avoid overfitting.

\begin{figure}
\includegraphics[width=0.9\linewidth]{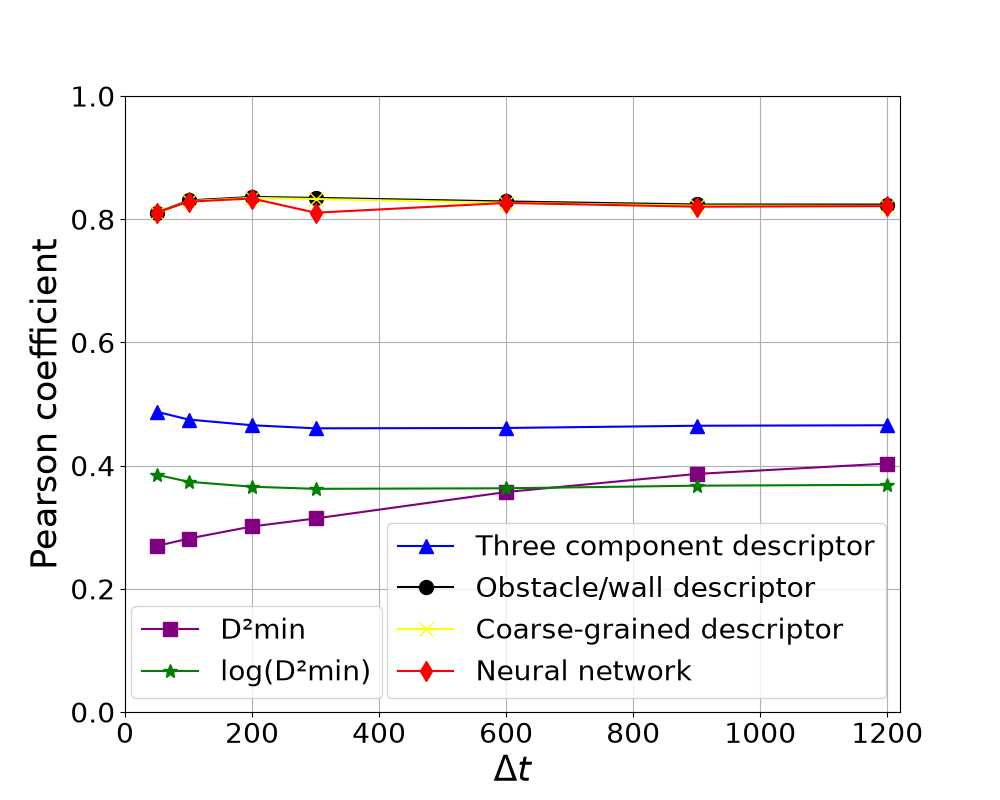}
\caption{Pearson correlation coefficient $\rho_{\hat Y,Y}$ measured on the test dataset, quantifying the performance of the machine-learning regression task as a function of the prediction timescale $\Delta t$. 
Successive curves show the improvement obtained by increasing the complexity of the machine-learning model.
}
\label{fig:performance_Dmin2}
\end{figure}

In Fig.~\ref{fig:performance_Dmin2}, we show the test-set performance of the linear prediction model described above. The performance is quantified by the Pearson correlation coefficient and plotted as a function of the prediction timescale $\Delta t$. The results obtained using the one-component BP descriptors and $D_{\rm min}^2$ as the target variable are shown by the purple squares. The performance increases slightly with $\Delta t$, but the overall value remains relatively small, around $0.4$, indicating limited predictive performance. Nevertheless, this simple setting provides a useful baseline for the rest of the paper, where we examine several possible improvements.

First, plastic activity is strongly localized around the obstacle in the ground-truth dataset, while most particles in the system remain nearly immobile. As a result, large values of $D_{\rm min}^2$ are rare, which may make the regression task difficult. To reduce the effect of this strong imbalance in the target variable, we also consider a logarithmic transformation, $\log D_{\rm min}^2$. This transformation slightly improves the performance, especially at shorter timescales where plastic activity is weak, as shown by the green stars in Fig.~\ref{fig:performance_Dmin2}. However, the overall performance remains essentially unchanged.

In addition to modifying the target variable $Y_i$, we also modify the input feature vector ${\bf X}_i$. Since the system is polydisperse, the standard one-component BP descriptors, which ignore particle-size information, may not fully capture the relevant local structure. We therefore divide the particles into small (S), medium (M), and large (L) species, chosen so that the three species have equal concentrations, and construct the corresponding three-component BP descriptors described in Appendix~\ref{sec:BP}. This procedure increases the number of input features and incorporates species-dependent structural information. We find that this modification modestly improves the performance, as shown by the blue triangles in Fig.~\ref{fig:performance_Dmin2}, with the Pearson correlation coefficient reaching values around $0.5$. However, the prediction performance still remains rather limited.

We notice that the machine-learning prediction does not capture plastic activity near the obstacle very well. This may be because the BP descriptors are constructed to be translationally and rotationally invariant~\cite{behler2007generalized}, whereas the presence of the obstacle explicitly breaks these symmetries. This symmetry breaking is physically important in the present system and may therefore reduce the predictive performance of the standard BP descriptors.

To address this issue, we introduce additional structural features that explicitly encode the presence of the obstacle and the walls. For each particle $i$, we define obstacle-related features as
\begin{align}
    \cos(m\theta_i)
    \exp\left(
    -\frac{d_{{\rm obs},i}}{\ell_{\rm obs}}
    \right), \\
    \sin(n\theta_i)
    \exp\left(
    -\frac{d_{{\rm obs},i}}{\ell_{\rm obs}}
    \right),
\end{align}
where $d_{{\rm obs},i}$ is the distance between particle $i$ and the center of the obstacle, and $\theta_i$ is the angle of particle $i$ measured with respect to the $x$ axis from the center of the obstacle. We use decay length scales $\ell_{\rm obs}=0.5,\ 1.0,\ 2.0,\ 5.0$
and angular modes
$m=0,\ 1,\ 2,\ 3,\ 4$ and $n=1,\ 2,\ 3,\ 4$.
These features encode both the distance from the obstacle and the angular position around it.

Similarly, we introduce wall-related features as
\begin{equation}
    \exp\left(
    -\frac{d_{{\rm wall},i}}{\ell_{\rm wall}}
    \right),
\end{equation}
where $d_{{\rm wall},i}$ is the distance from particle $i$ to the nearest wall. We use the decay length scales
$\ell_{\rm wall}=0.5,\ 1.0,\ 2.0,\ 5.0,  \ 10.0$.
These additional features explicitly provide information about the symmetry-breaking boundaries of the system.

In Fig.~\ref{fig:performance_Dmin2}, we show the performance obtained after adding the obstacle and wall features on top of the previous descriptors. We find that the Pearson correlation coefficient increases dramatically. This improvement indicates that the additional features allow the machine-learning model to focus more effectively on the region near the obstacle, where plastic activity is most likely to occur.

However, this result should be interpreted with some caution. The improvement does not necessarily mean that the model has learned the detailed heterogeneous structure of plastic activity. Instead, it may simply reflect the fact that the model has learned that particles close to the obstacle are more likely to be mobile.

To further improve the prediction, we also consider coarse-grained descriptors, following the approach introduced by the group of Filion~\cite{boattini2021averaging}, which was shown to improve the prediction of dynamics in glass-forming liquids significantly. 
The numerical procedure for coarse-graining, implemented as a local averaging operation, is described in Appendix~\ref{sec:BP}.
The coarse-graining procedure is applied only to the BP descriptors, and not to the obstacle and wall features.

In Fig.~\ref{fig:performance_Dmin2}, we find that this coarse-graining procedure does not further improve the prediction performance in the present foam-flow system, unlike for glass-forming liquids data.

\begin{figure*}[htbp]
    \centering
    \begin{subfigure}[t]{0.32\linewidth}
        \captionsetup{justification=Justified, singlelinecheck=false, position=above}  
        \caption{Ground truth}
        \includegraphics[width=\linewidth]{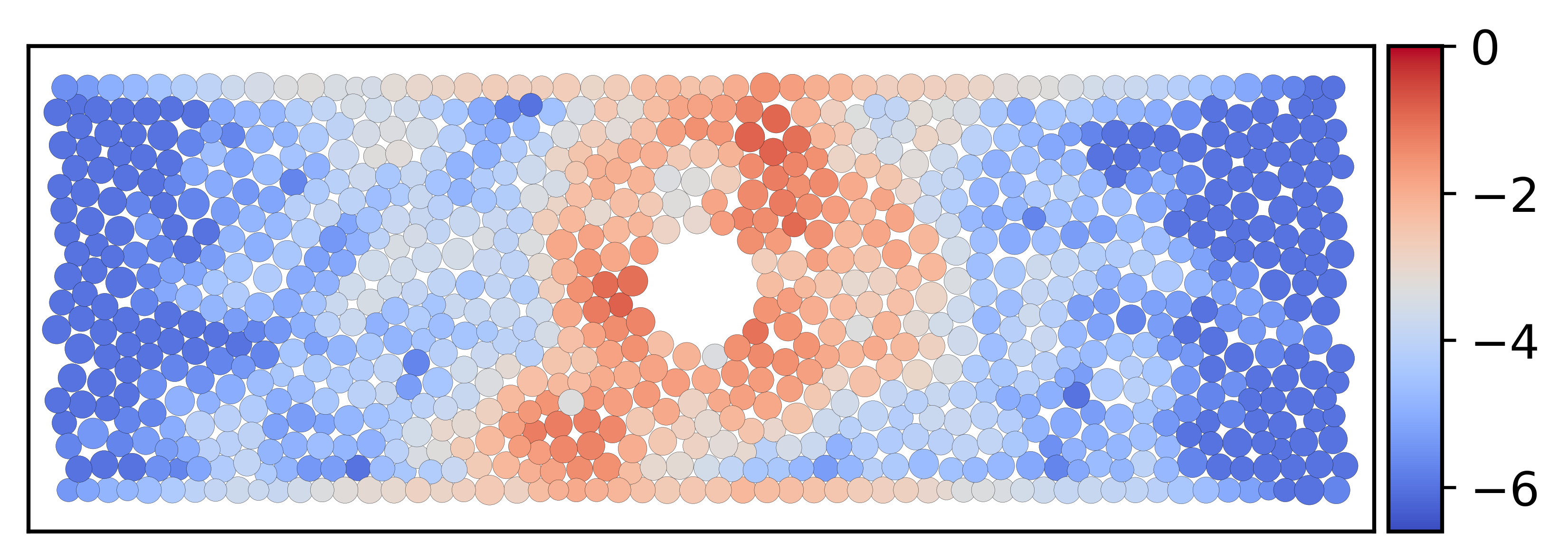}
        \label{fig:delta10a}
    \end{subfigure}
    \hfill
    \begin{subfigure}[t]{0.32\linewidth}
        \captionsetup{justification=Justified, singlelinecheck=false, position=above}
        \caption{Linear regression prediction}
        \includegraphics[width=\linewidth]{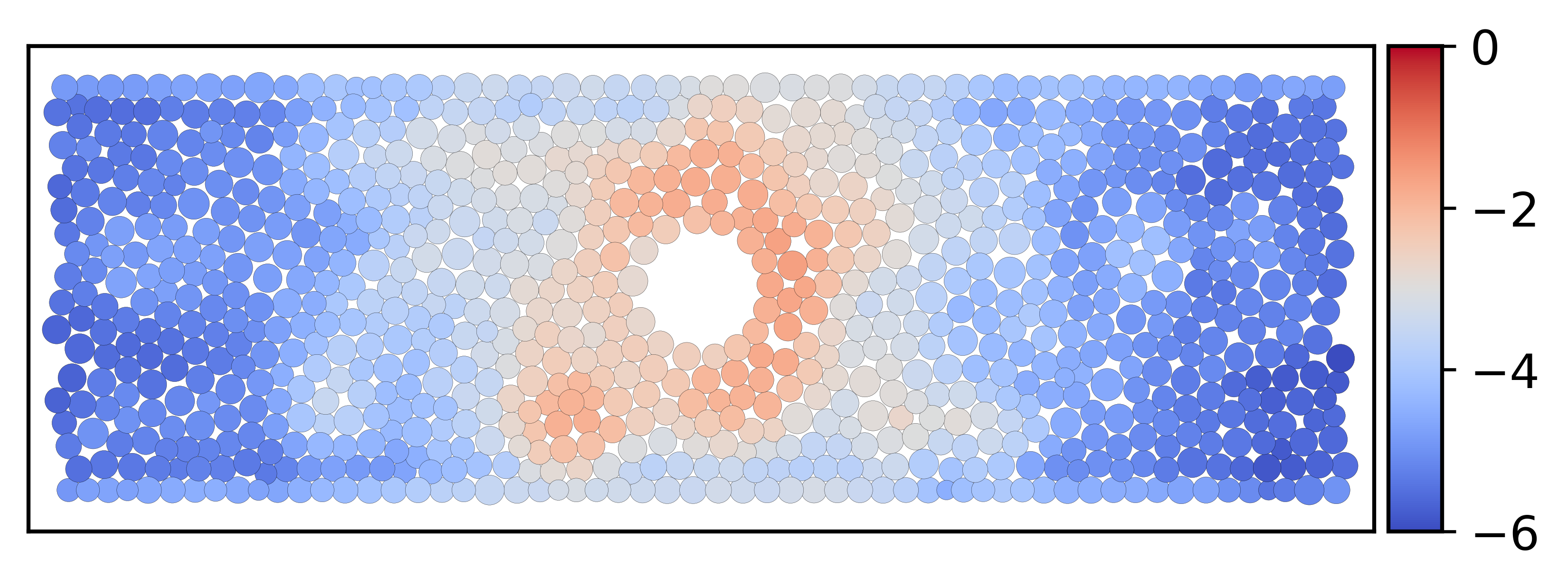}
        \label{fig:delta10b}
    \end{subfigure}
    \hfill
    \begin{subfigure}[t]{0.32\linewidth}
        \captionsetup{justification=Justified, singlelinecheck=false, position=above}
        \caption{Neural network prediction}
        \includegraphics[width=\linewidth]{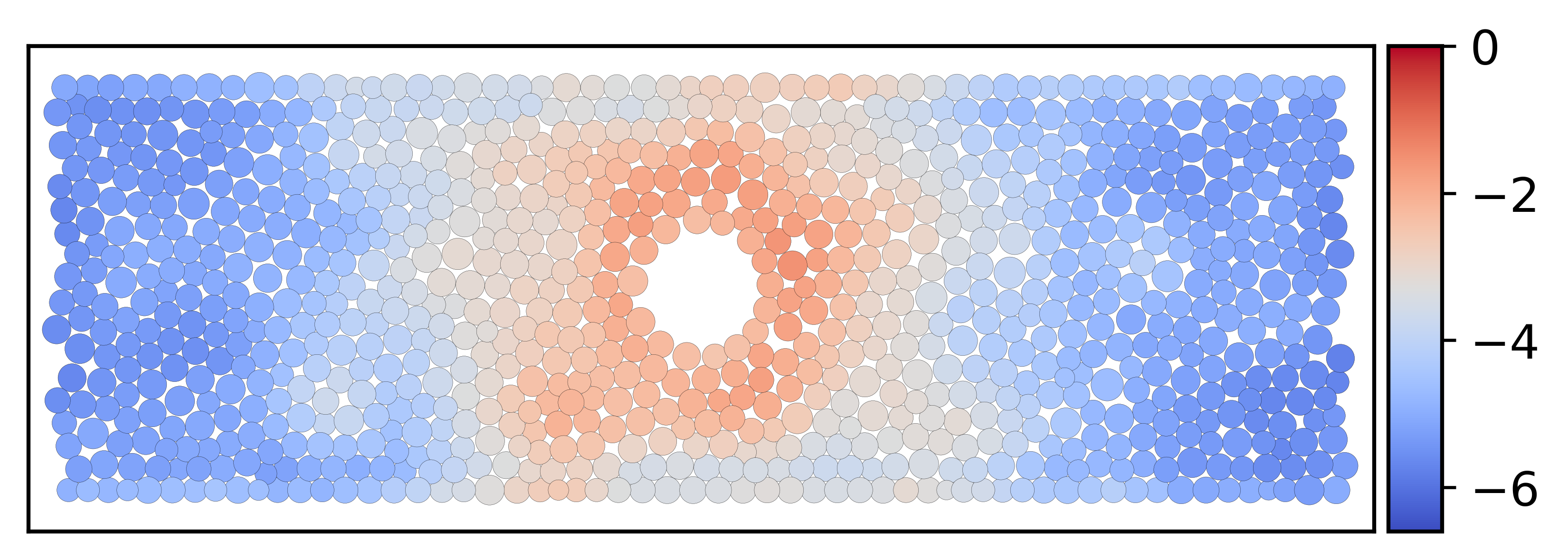}
        \label{fig:delta10c}
    \end{subfigure}
    \vspace{3mm}
    \begin{subfigure}[t]{0.32\linewidth}
        \captionsetup{justification=Justified, singlelinecheck=false, position=above}
        \caption{Ground truth}
        \includegraphics[width=\linewidth]{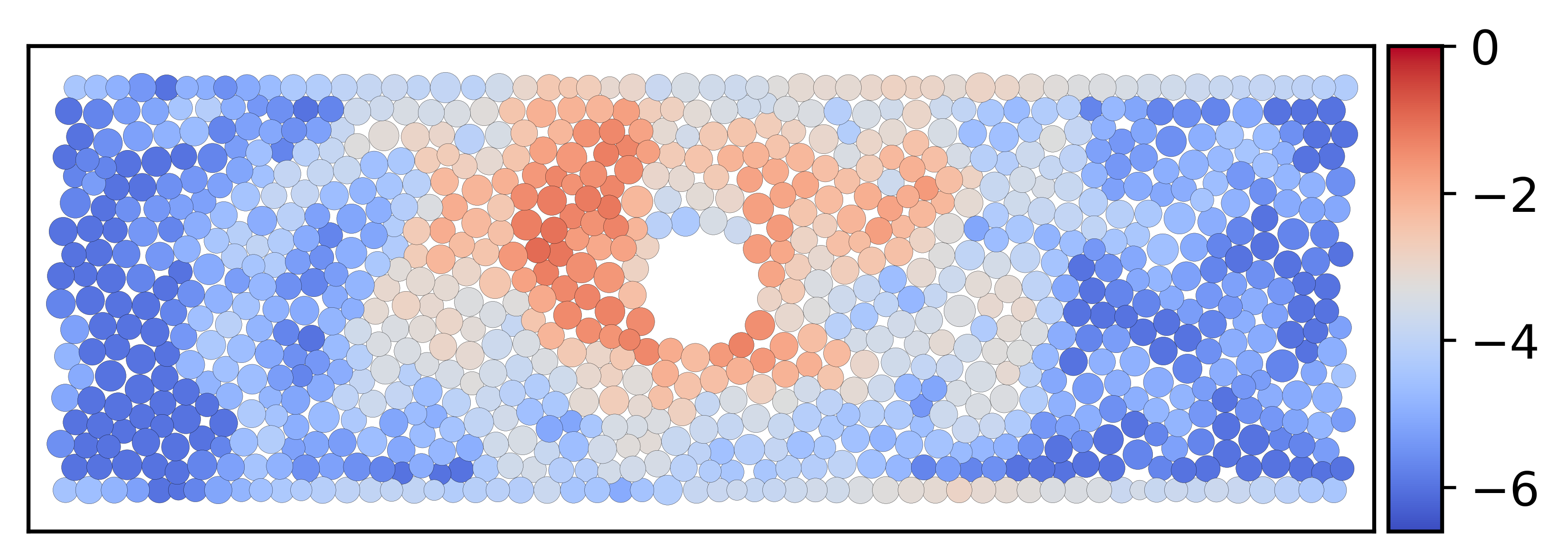}
        \label{fig:delta30a}
    \end{subfigure}
    \hfill
    \begin{subfigure}[t]{0.32\linewidth}
        \captionsetup{justification=Justified, singlelinecheck=false, position=above}
        \caption{Linear regression prediction}
        \includegraphics[width=\linewidth]{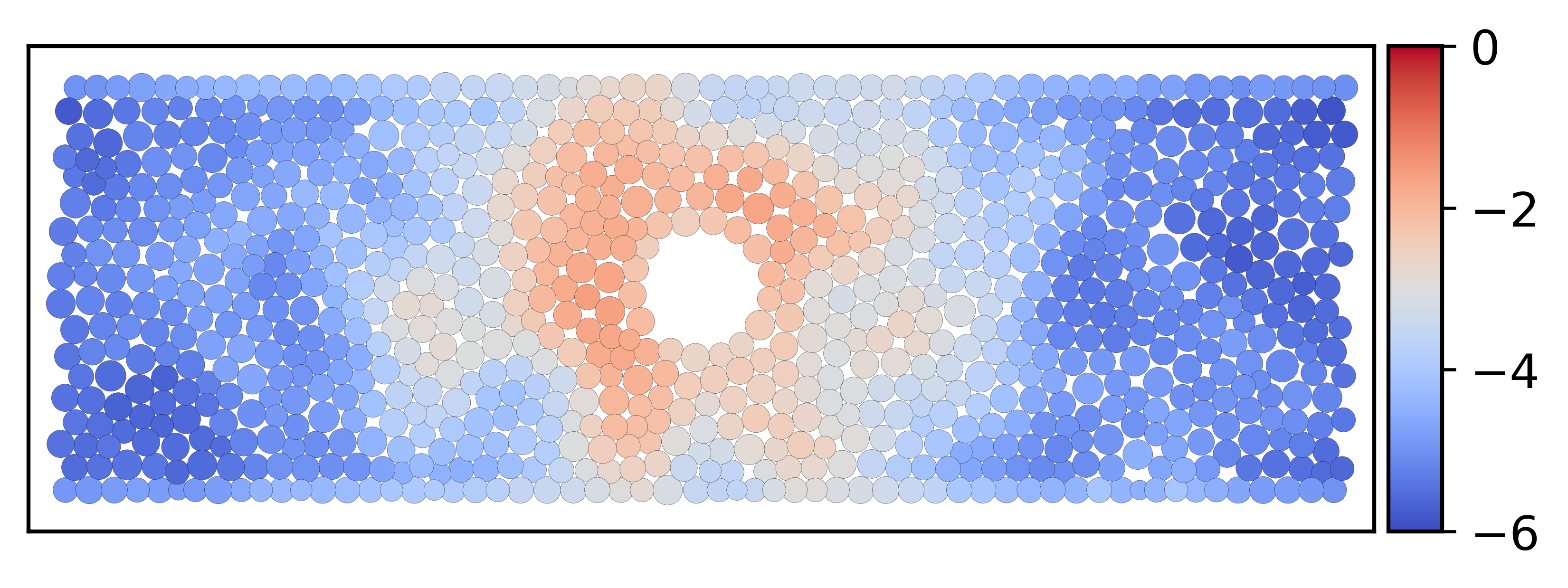}
        \label{fig:delta30b}
    \end{subfigure}
    \hfill
    \begin{subfigure}[t]{0.32\linewidth}
        \captionsetup{justification=Justified, singlelinecheck=false, position=above}
        \caption{Neural network prediction}
        \includegraphics[width=\linewidth]{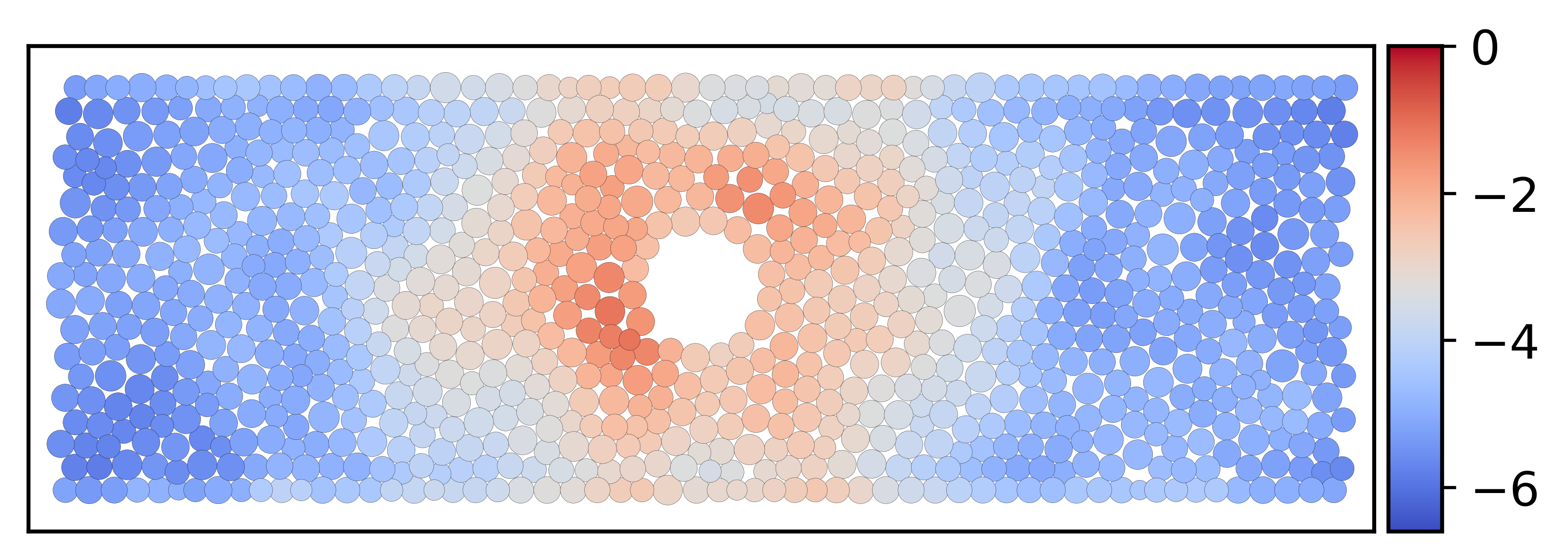}
        \label{fig:delta30c}
    \end{subfigure}
    \vspace{3mm}
    \begin{subfigure}[t]{0.32\linewidth}
        \captionsetup{justification=Justified, singlelinecheck=false, position=above}
        \caption{Ground truth}
        \includegraphics[width=\linewidth]{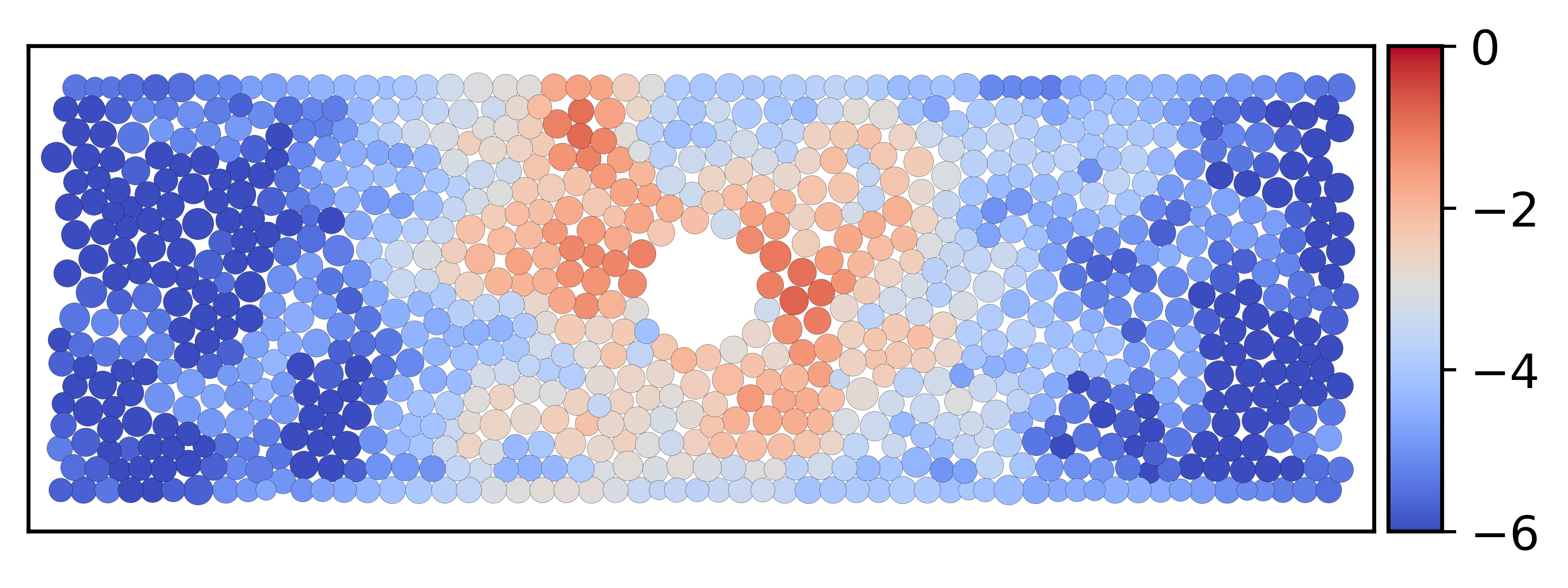}
        \label{fig:delta120a}
    \end{subfigure}
    \hfill
    \begin{subfigure}[t]{0.32\linewidth}
        \captionsetup{justification=Justified, singlelinecheck=false, position=above}
        \caption{Linear regression prediction}
        \includegraphics[width=\linewidth]{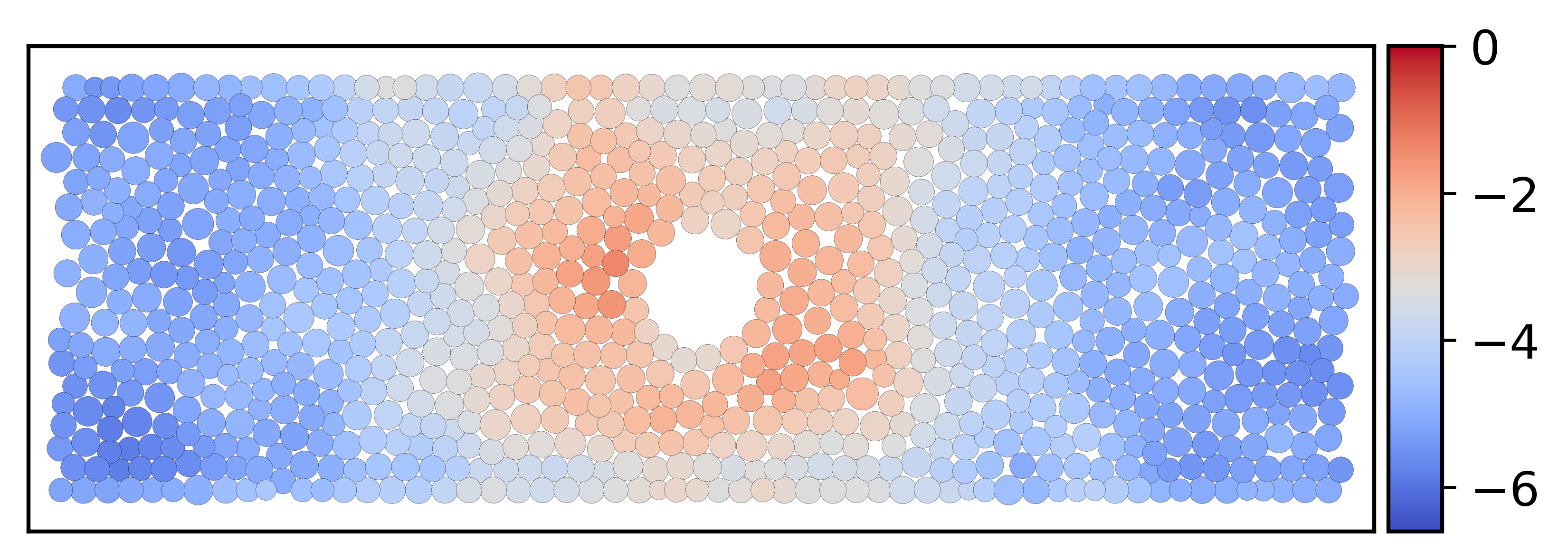}
        \label{fig:delta120b}
    \end{subfigure}
    \hfill
    \begin{subfigure}[t]{0.32\linewidth}
        \captionsetup{justification=Justified, singlelinecheck=false, position=above}
        \caption{Neural network prediction}
        \includegraphics[width=\linewidth]{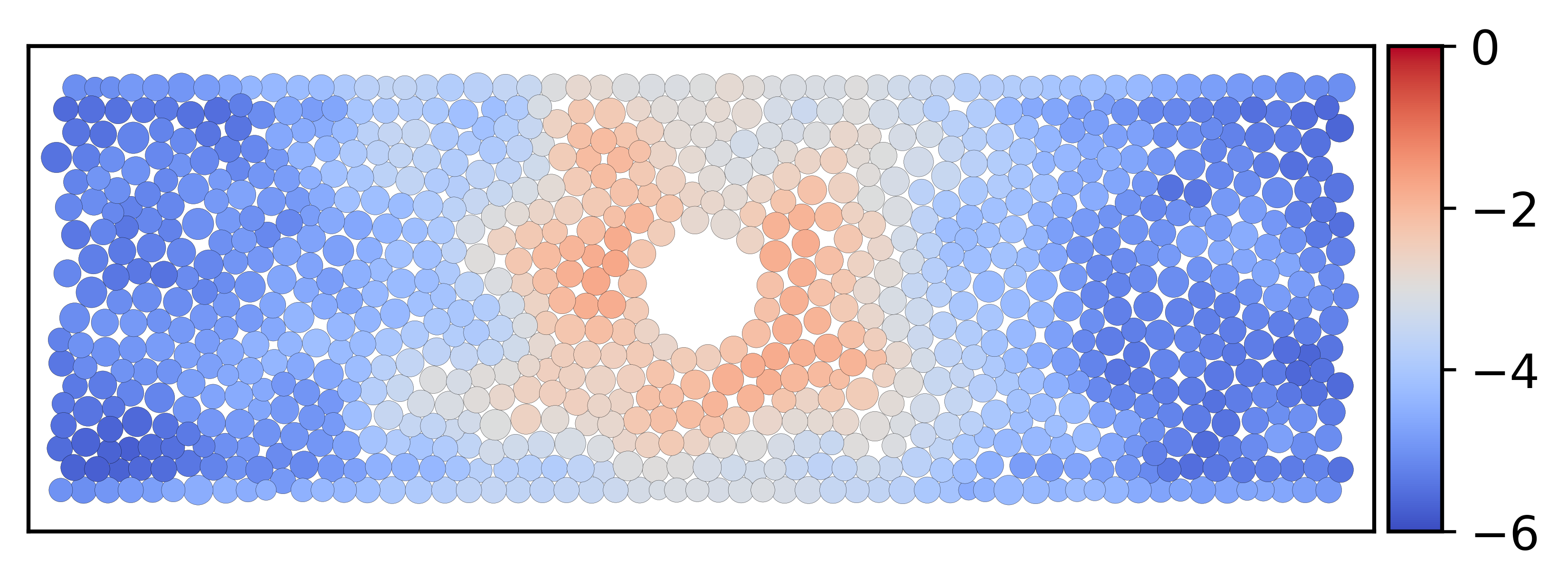}
        \label{fig:delta120c}
    \end{subfigure}
\caption{Prediction of $\log D_{\rm min}^2$ for $\Delta t=300$ using linear regression and a neural network, compared with the ground truth in the test dataset for three representative configurations. 
(a,d,g) Ground truth, (b,e,h) corresponding linear-regression predictions, and (c,f,i) corresponding neural-network predictions.}
    \label{fig:ground_truth_vs_prediction_Dmin2}
\end{figure*}

The best linear-regression prediction considered so far is visualized in Fig.~\ref{fig:ground_truth_vs_prediction_Dmin2}, where we compare the ground truth in the test dataset with the corresponding prediction.
We find that, although linear regression can predict enhanced plastic activity near the obstacle, it still fails to capture the detailed heterogeneous pattern observed in the ground truth.

What prevents us from predicting the detailed heterogeneous dynamics? One natural possibility is that the linear model in Eq.~\eqref{eq:linear_model} is too simple to capture plastic activity, and that nonlinear modeling is necessary. To test this hypothesis, we employ a multilayer perceptron (MLP) neural network, following Ref.~\cite{alkemade2022comparing}.

We consider an MLP, which is a fully connected feedforward neural network composed of an input layer, hidden layers, and an output layer. The input layer receives the feature vector ${\bf X}_i$, while the output layer predicts the target variable, $Y_i=\log D_{\rm min}^2$. Between the input and output layers, the network contains hidden layers. The number of hidden layers and the number of neurons in each layer are hyperparameters. Neurons in adjacent layers are connected by trainable weights, and each neuron applies a nonlinear activation function before transmitting information to the next layer. We use the rectified linear unit (ReLU) activation function, defined as $\operatorname{ReLU}(z) = \max(0,z)$,
which leaves positive inputs unchanged and sets negative inputs to zero, thereby introducing nonlinearity into the network.

The network is trained by minimizing a mean-squared-error loss of the same form as Eq.~\eqref{eq:loss_MSE}, but with a much larger number of trainable weight parameters due to the presence of multiple neurons and layers. 
We optimize the network using the Adam stochastic-gradient-based optimizer with a batch size of 50~\cite{kingma2014adam,alkemade2022comparing}.

We consider three architectures: $(8)$, $(16)$, and $(8,8)$. Here, $(8)$ and $(16)$ denote networks with a single hidden layer containing 8 and 16 neurons, respectively, while $(8,8)$ denotes a network with two hidden layers, each containing 8 neurons.
The three architectures yield essentially the same performance; therefore, in this paper we report only the results obtained with the $(8,8)$ model.
We confirmed that overfitting is negligible from the training curves obtained by varying the number of configurations in the training dataset, as shown in Appendix~\ref{sec:training_curves}.
%We therefore did not introduce an explicit $L_2$ penalty and set $\alpha=0$.

As shown in Fig.~\ref{fig:performance_Dmin2}, the prediction performance of the neural network does not improve compared with the linear model. 
Visual inspection in Fig.~\ref{fig:ground_truth_vs_prediction_Dmin2} suggests that the neural network may slightly improve the prediction of heterogeneous patterns compared with linear regression. 
However, the overall performance remains largely unchanged. 
This result indicates that introducing nonlinearity with respect to the current feature vector ${\bf X}_i$ is not sufficient to substantially improve the prediction.

\subsection{Classification of neighbor change events}

\begin{figure}
\includegraphics[width=0.9\linewidth]{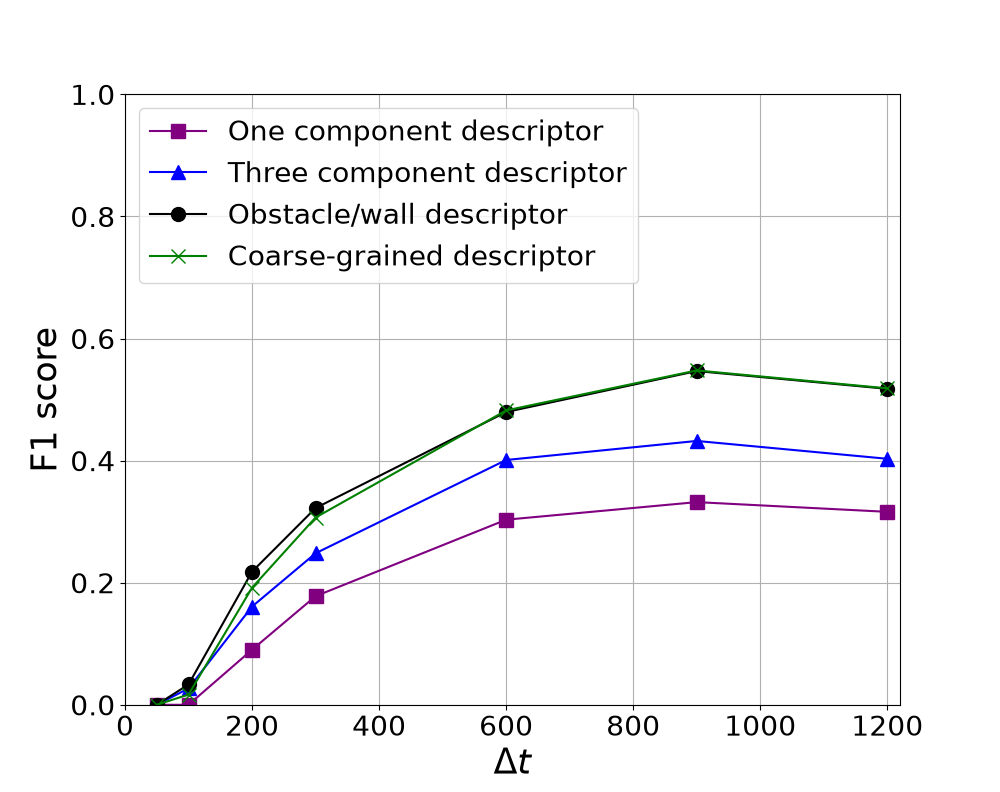}
\caption{F1 score for the prediction of neighbor change events as a function of the prediction timescale $\Delta t$, with successive improvements obtained by increasing the complexity of the machine-learning model.}
\label{fig:performance_T1}
\end{figure}

\begin{figure*}[htbp]
    \centering
    \begin{subfigure}[t]{0.32\linewidth}
        \captionsetup{justification=Justified, singlelinecheck=false, position=above}  
        \caption{Ground truth}
        \includegraphics[width=\linewidth]{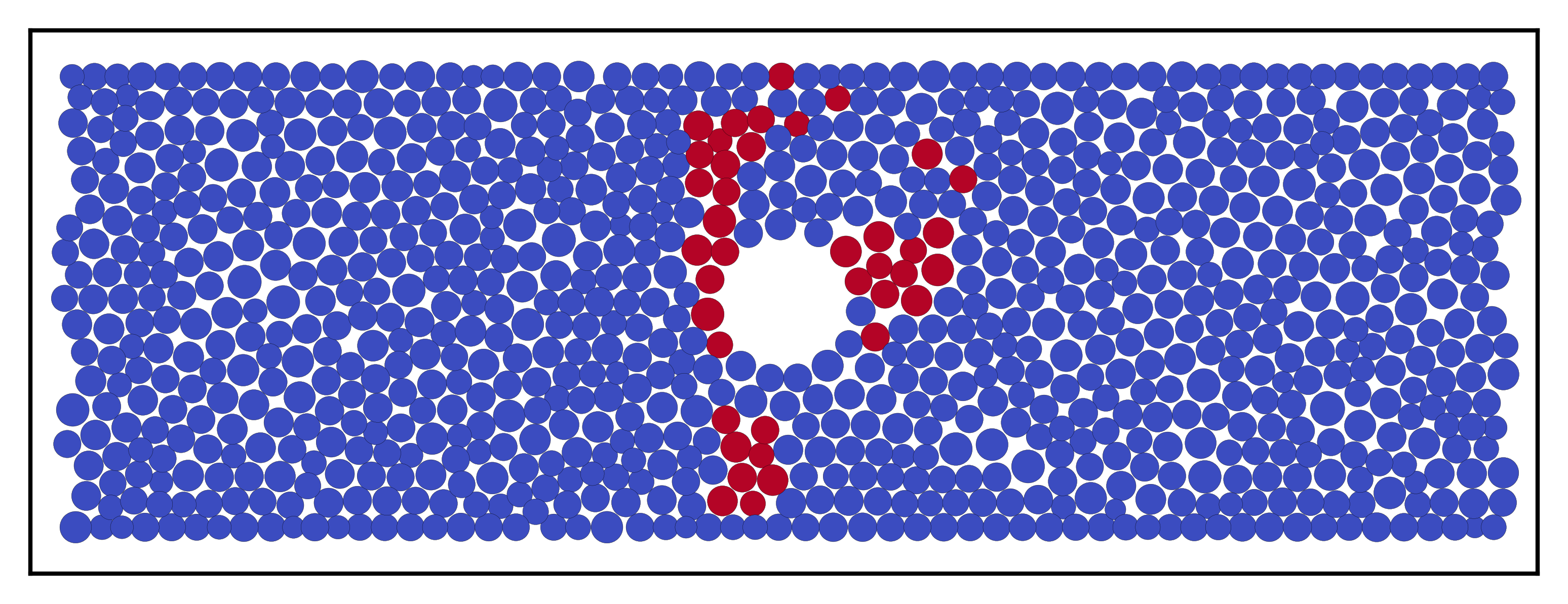}
        \label{fig:delta10a}
    \end{subfigure}
    \hfill
    \begin{subfigure}[t]{0.32\linewidth}
        \captionsetup{justification=Justified, singlelinecheck=false, position=above}
        \caption{Probability prediction}
        \includegraphics[width=\linewidth]{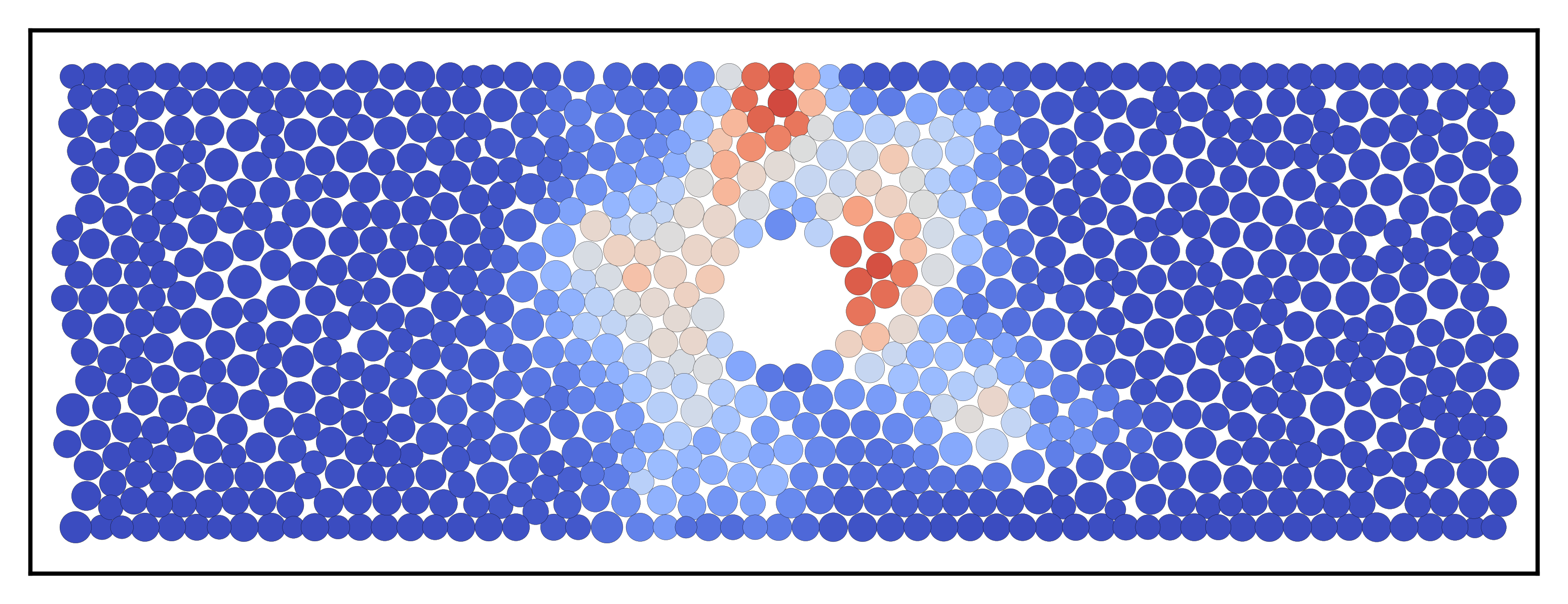}
        \label{fig:delta10b}
    \end{subfigure}
    \hfill
    \begin{subfigure}[t]{0.32\linewidth}
        \captionsetup{justification=Justified, singlelinecheck=false, position=above}
        \caption{Binary prediction}
        \includegraphics[width=\linewidth]{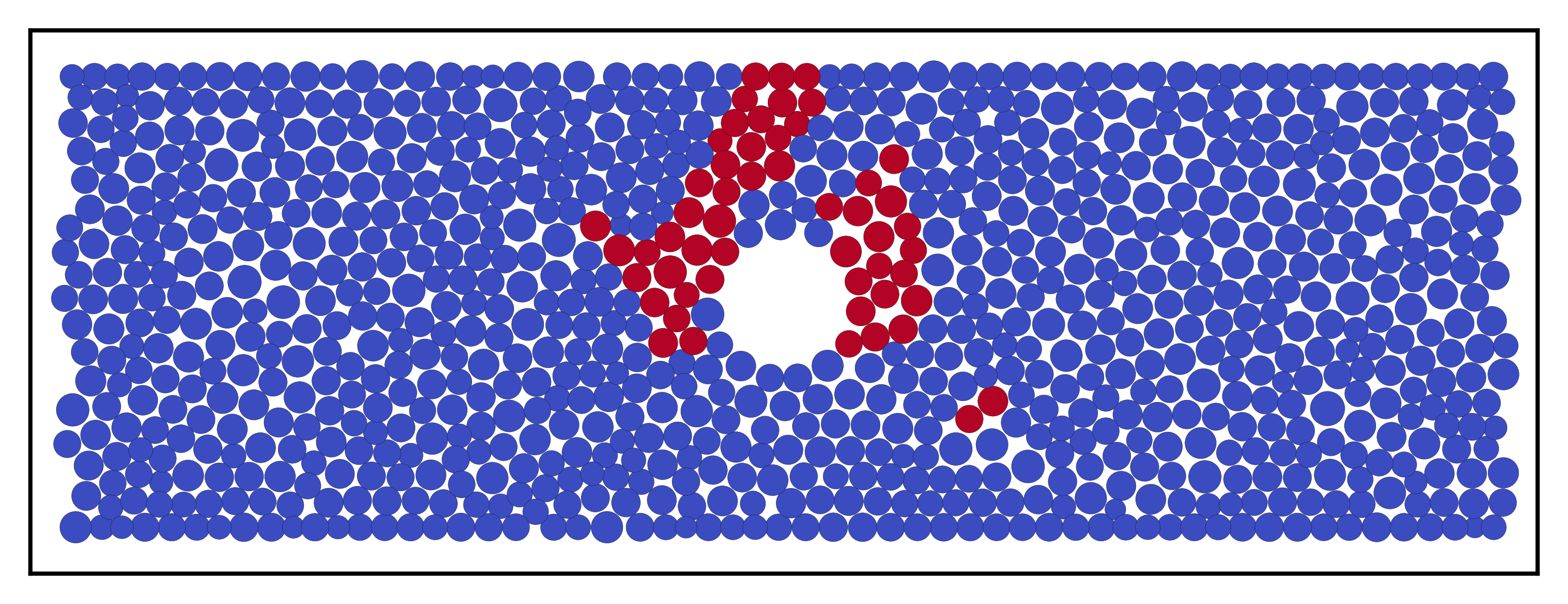}
        \label{fig:delta10c}
    \end{subfigure}
    \vspace{3mm}
    \begin{subfigure}[t]{0.32\linewidth}
        \captionsetup{justification=Justified, singlelinecheck=false, position=above}
        \caption{Ground truth}
        \includegraphics[width=\linewidth]{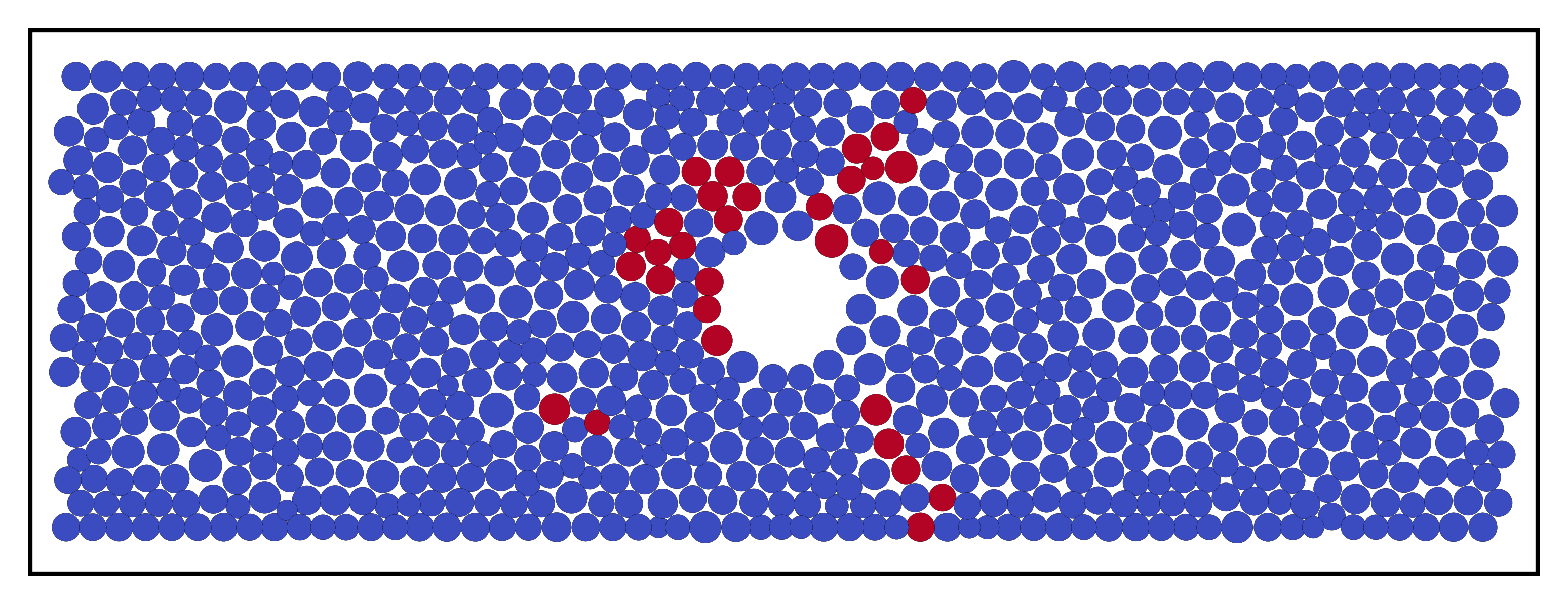}
        \label{fig:delta30a}
    \end{subfigure}
    \hfill
    \begin{subfigure}[t]{0.32\linewidth}
        \captionsetup{justification=Justified, singlelinecheck=false, position=above}
        \caption{Probability prediction}
        \includegraphics[width=\linewidth]{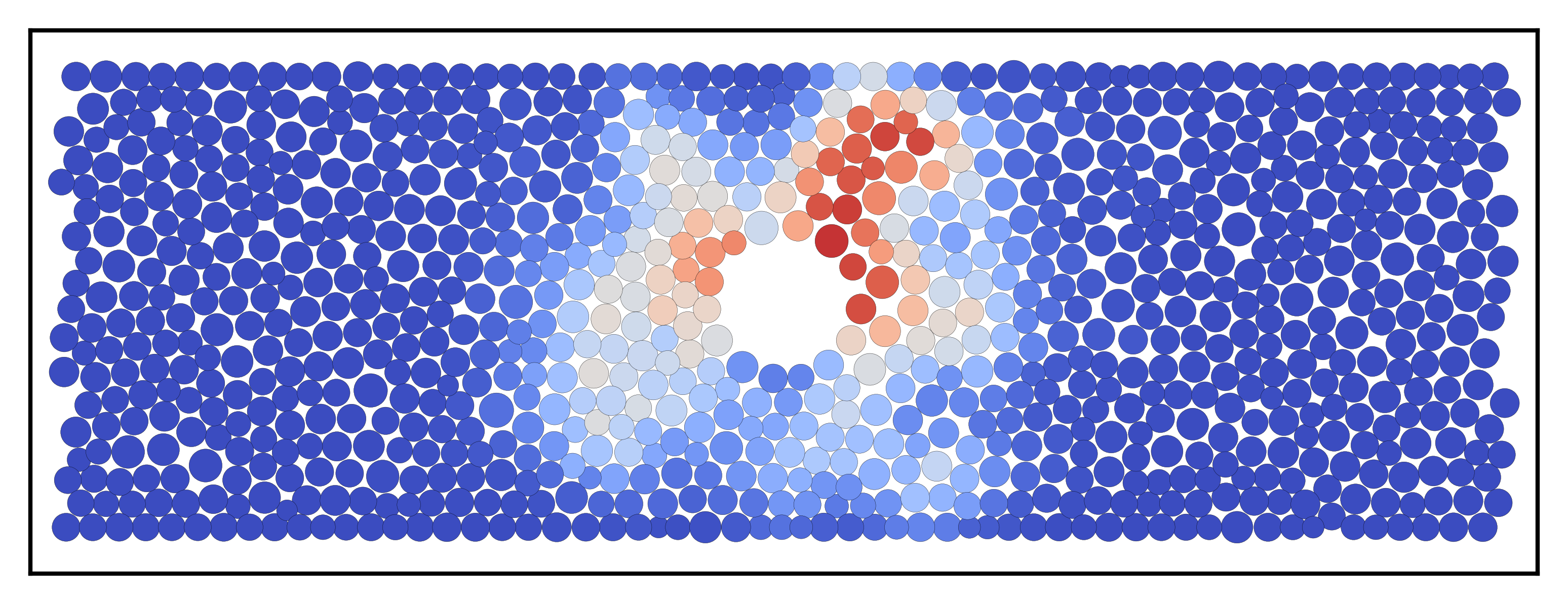}
        \label{fig:delta30b}
    \end{subfigure}
    \hfill
    \begin{subfigure}[t]{0.32\linewidth}
        \captionsetup{justification=Justified, singlelinecheck=false, position=above}
        \caption{Binary prediction}
        \includegraphics[width=\linewidth]{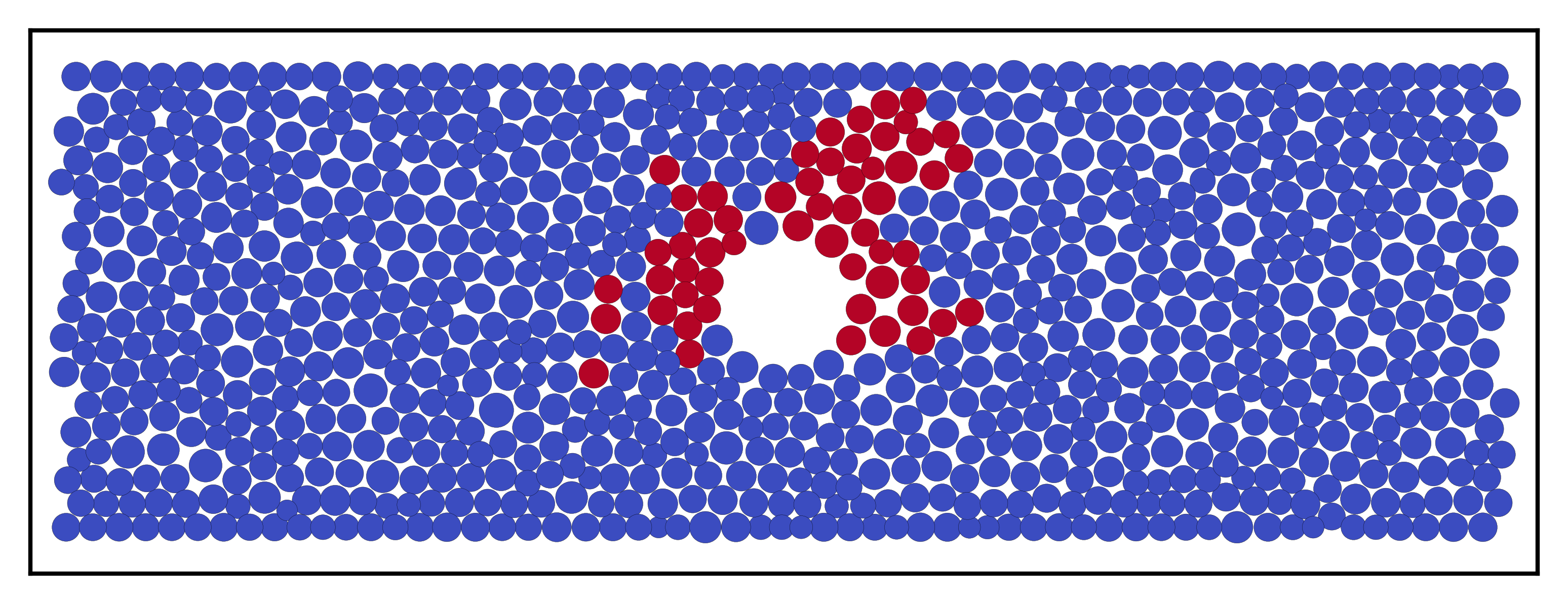}
        \label{fig:delta30c}
    \end{subfigure}
    \vspace{3mm}
    \begin{subfigure}[t]{0.32\linewidth}
        \captionsetup{justification=Justified, singlelinecheck=false, position=above}
        \caption{Ground truth}
        \includegraphics[width=\linewidth]{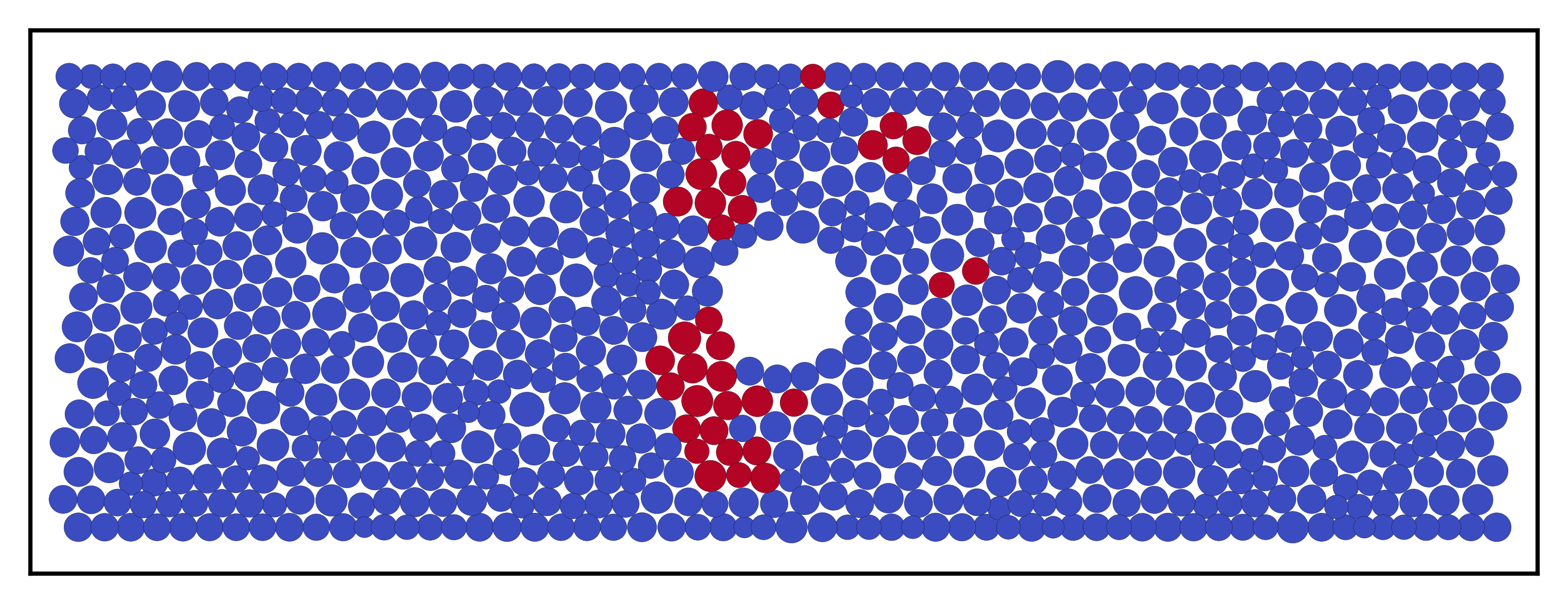}
        \label{fig:delta120a}
    \end{subfigure}
    \hfill
    \begin{subfigure}[t]{0.32\linewidth}
        \captionsetup{justification=Justified, singlelinecheck=false, position=above}
        \caption{Probability prediction}
        \includegraphics[width=\linewidth]{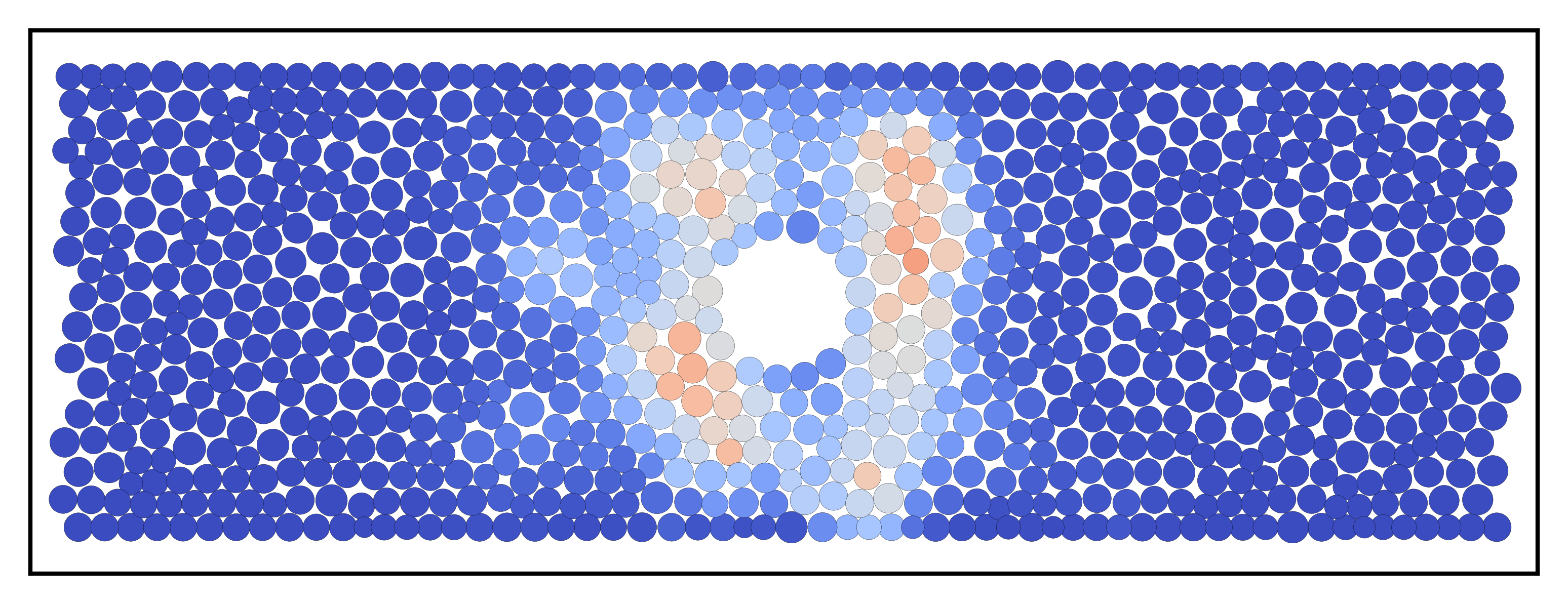}
        \label{fig:delta120b}
    \end{subfigure}
    \hfill
    \begin{subfigure}[t]{0.32\linewidth}
        \captionsetup{justification=Justified, singlelinecheck=false, position=above}
        \caption{Binary prediction}
        \includegraphics[width=\linewidth]{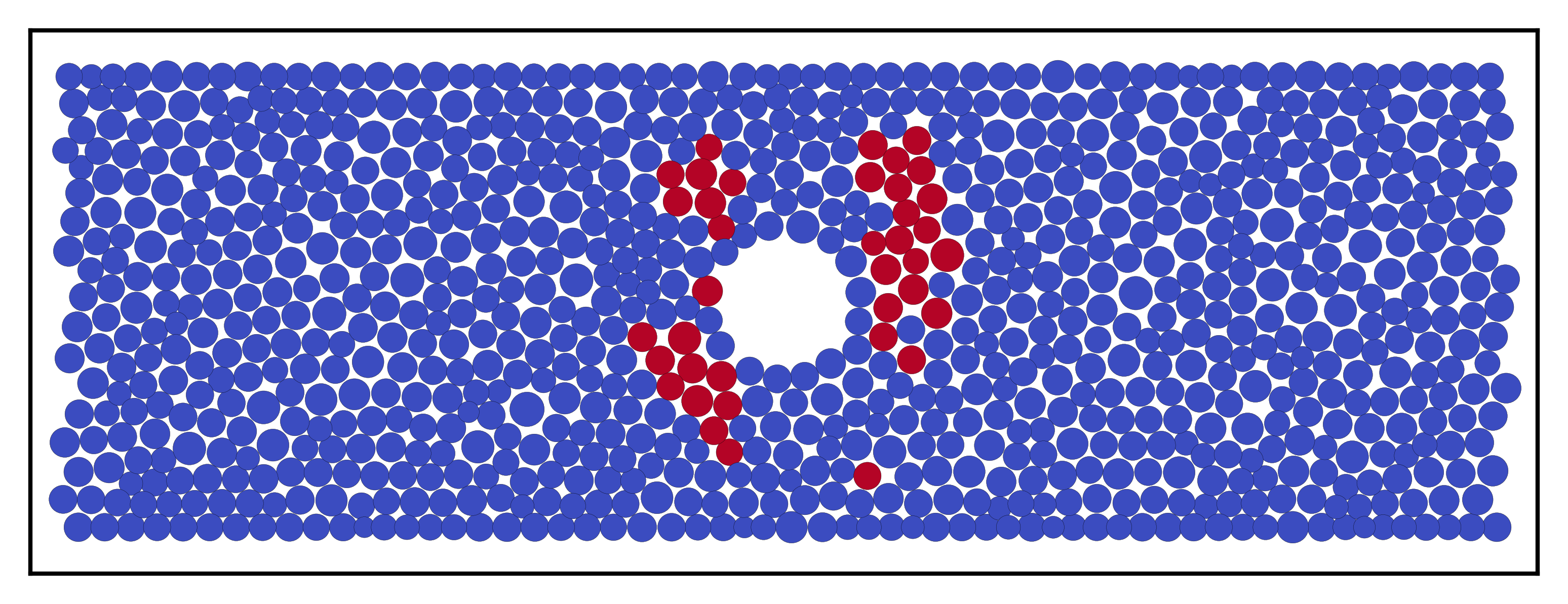}
        \label{fig:delta120c}
    \end{subfigure}
    \caption{Prediction of neighbor change events for $\Delta t=300$ using logistic regression, compared with the ground truth in the test dataset for three representative configurations. 
(a,d,g) Ground truth, (b,e,h) predicted probability, and (c,f,i) corresponding binary prediction.}
    \label{fig:ground_truth_vs_prediction_T1}
\end{figure*}

We next treat the prediction of plasticity as a binary classification problem for neighbor change events. The target label is now given by the neighbor change indicator in Eq.~(\ref{eq:T1_indicator}), $Y_i=I_i$, while the input structural feature vector ${\bf X}_i$ is the same as that used in the regression problem above. 

We employ a binary logistic regression scheme.
We consider the probabilistic model
\begin{equation}
    P_{\bf w}(Y_i=1|{\bf X}_i)
    =
    \frac{1}{1+\exp[-{\bf w}^{\rm T}{\bf X}_i]} ,
\end{equation}
where $P_{\bf w}(Y_i=1|{\bf X}_i)$ denotes the probability that particle $i$ undergoes a neighbor change event, given the feature vector ${\bf X}_i$. The probability of no event is instead given by
$P_{\bf w}(Y_i=0|{\bf X}_i) = 1-P_{\bf w}(Y_i=1|{\bf X}_i)$.

The weight vector is determined by minimizing the following loss function,
\begin{eqnarray}
\mathcal{L}({\bf w})
    &=&
    -  \sum_{i=1}^{N_{\rm train}}
    \Big[
    c_1 Y_i \ln P_{\bf w}(Y_i=1|{\bf X}_i) \nonumber \\
    &\qquad&  \qquad +
    c_0(1-Y_i)\ln P_{\bf w}(Y_i=0|{\bf X}_i)
    \Big] ,
\end{eqnarray}
where $c_1$ and $c_0$ are class weights. These class weights (hyperparameters) are introduced to account for the strong class imbalance in the dataset, where the number of positive samples, $Y_i=1$, is much smaller than the number of negative samples, $Y_i=0$. We set $c_0=1$ and choose $c_1=6$ to compensate for this imbalance.

The prediction performance is evaluated using the F1 score~\cite{sokolova2009systematic}, which is particularly useful for imbalanced classification problems such as the present plasticity prediction task~\cite{rocks2021learning}. A detailed introduction to the F1 score is given in Appendix~\ref{sec:class_performance}.
We also checked that overfitting is negligible by examining the training curves as a function of the number of training configurations, as shown in Appendix~\ref{sec:training_curves}.
Thus, we do not introduce an explicit $L_2$ penalty.

In Fig.~\ref{fig:performance_T1}, we show the performance of the neighbor change event prediction as a function of the prediction timescale $\Delta t$. As in the regression approach, we start from the simplest setting using one-component BP descriptors, shown by the purple squares. The F1 score gradually increases, reaches a maximum of approximately $0.3$ around $\Delta t \approx 600-1000$, and then decreases at longer times, suggesting that predictability is highest at an intermediate timescale.

We then increase the complexity of the model step by step: three-component BP descriptors, shown by the blue triangles; the addition of obstacle and wall descriptors, shown by the black circles; and coarse-grained BP descriptors, shown by the green crosses. Consistent with the regression task, the obstacle and wall descriptors lead to the largest improvement in performance. In contrast, the three-component descriptors provide only a modest improvement, and the coarse-graining of BP descriptors does not further improve the performance.

We visually compare the ground truth and the prediction at $\Delta t=300$ in Fig.~\ref{fig:ground_truth_vs_prediction_T1}. The prediction is shown both as the probability map $P_{\bf w}(Y_i=1|{\bf X}_i)$ and as the corresponding binary prediction, where particles are shown in red when $P_{\bf w}(Y_i=1|{\bf X}_i)\geq 1/2$ and in blue otherwise. As in the $D_{\rm min}^2$ regression task, the prediction is concentrated around the obstacle. 
Although the machine-learning model captures the overall spatial pattern to some extent, it still fails to reproduce the detailed heterogeneous pattern observed in the ground truth.

\section{How Much Does Structure Determine Future Plasticity?}
\label{sec:perturbation}

In the previous section, we studied the machine-learning prediction of plasticity using both regression and classification approaches. Although our approach, in particular the introduction of obstacle and wall features that explicitly break translational and rotational symmetries, significantly improves the overall performance, the machine-learning prediction still fails to fully capture the detailed heterogeneous pattern of plastic activity around the obstacle.

Logically, there are two possible scenarios to understand this limitation. The first possibility is that the present machine-learning approach, based on BP descriptors, is not expressive enough to capture the relevant structural information~\cite{jung2025roadmap}. The second possibility is that amorphous plasticity in confined channel foam flow around an obstacle is intrinsically difficult to predict from a single static snapshot. In other words, the static structure may not encode enough information about the future dynamics~\cite{berthier2007structure}, and the subsequent plastic activity may instead be highly sensitive to small perturbations of the initial configuration.

To distinguish between these two scenarios, we perform the following perturbation analysis. Starting from a given configuration, we perturb all particle positions according to
\begin{equation}
    {\bf r}_i \to {\bf r}_i + \delta {\bf r}_i ,
\end{equation}
where $\delta {\bf r}_i$ is sampled uniformly from a square box of linear size $d$, similarly to a Monte Carlo displacement~\cite{berthier2007monte}. More precisely, each coordinate can variate from $-d/2$ to $d/2$. We choose the perturbations so that the center of mass of the system does not shift. We then perform molecular-dynamics simulations from the perturbed configurations, solving Eq.~(\ref{eq:overdamped_foam}). This procedure is repeated 10 times from the same initial configuration.

We observe that, when $d$ is small, all perturbed trajectories follow essentially the same dynamical path. In contrast, when $d$ is very large, the perturbed trajectories become completely different. We then compute the average of $\log D_{\rm min}^2$ over 10 perturbed trajectories. This averaged quantity is analogous to the dynamic propensity used in studies of glass-forming liquids~\cite{widmer2004reproducible}.

\begin{figure*}[htbp]
    \centering
        \begin{subfigure}{0.32\textwidth}
        \captionsetup{justification=Justified, singlelinecheck=false, position=above}
        \caption{One trajectory $\log D_{\rm min}^2$ with $d=0$}
        \includegraphics[width=\textwidth]{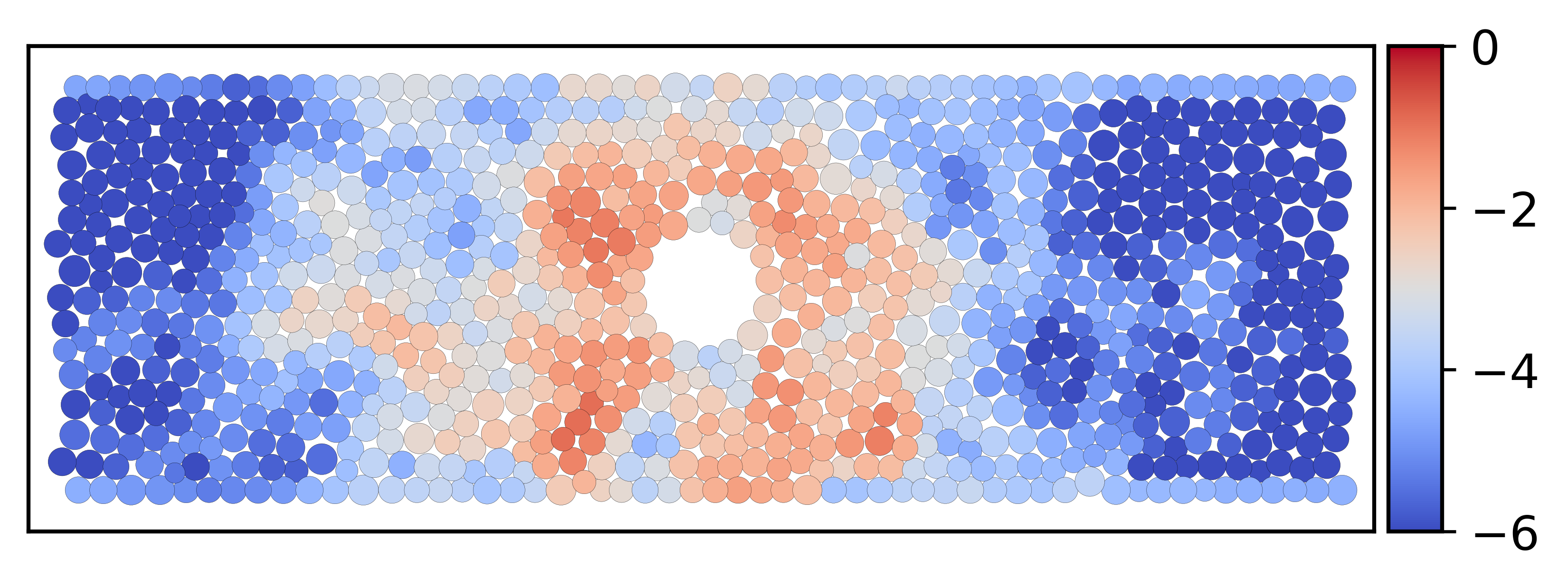}
        \label{fig:c}
    \end{subfigure}
    \hfill
    \begin{subfigure}{0.32\textwidth}
        \captionsetup{justification=Justified, singlelinecheck=false, position=above}
        \caption{Averaged $\log D_{\rm min}^2$ with $d=0.3$}
        \includegraphics[width=\textwidth]{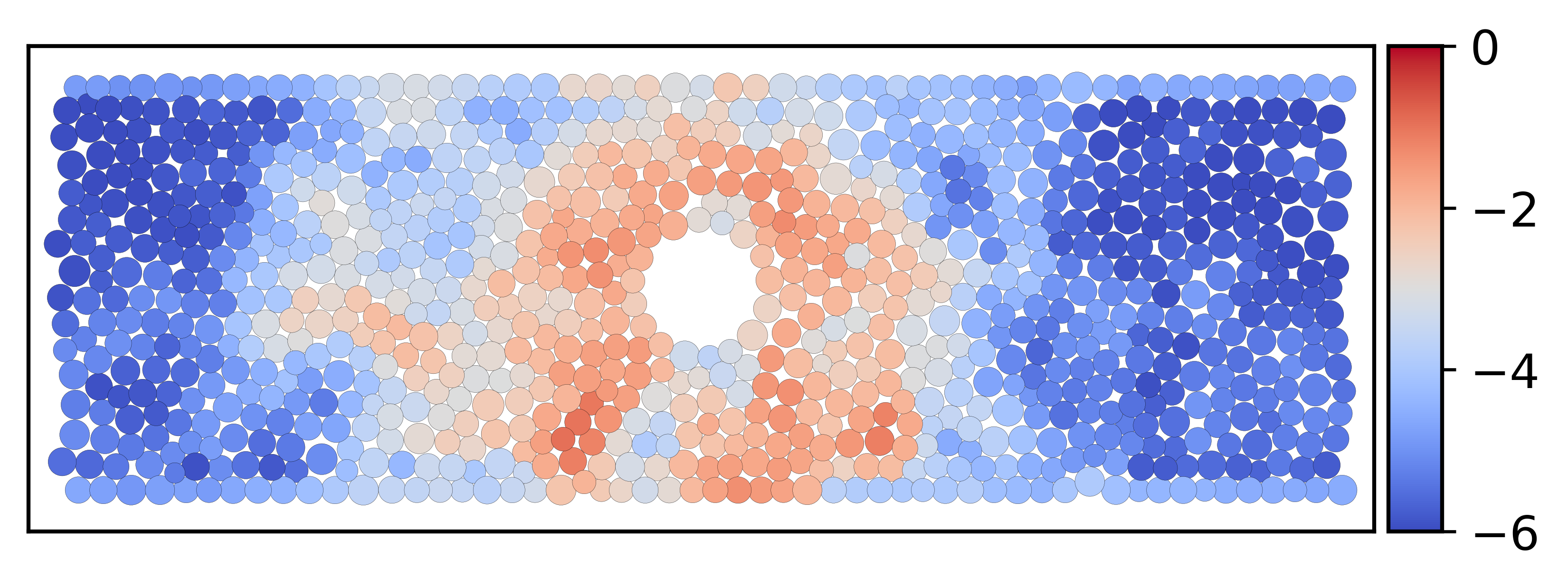}
        \label{fig:a}
    \end{subfigure}
    \hfill
    \begin{subfigure}{0.32\textwidth}
        \captionsetup{justification=Justified, singlelinecheck=false, position=above}
        \caption{Averaged $\log D_{\rm min}^2$ with $d=0.5$}
        \includegraphics[width=\textwidth]{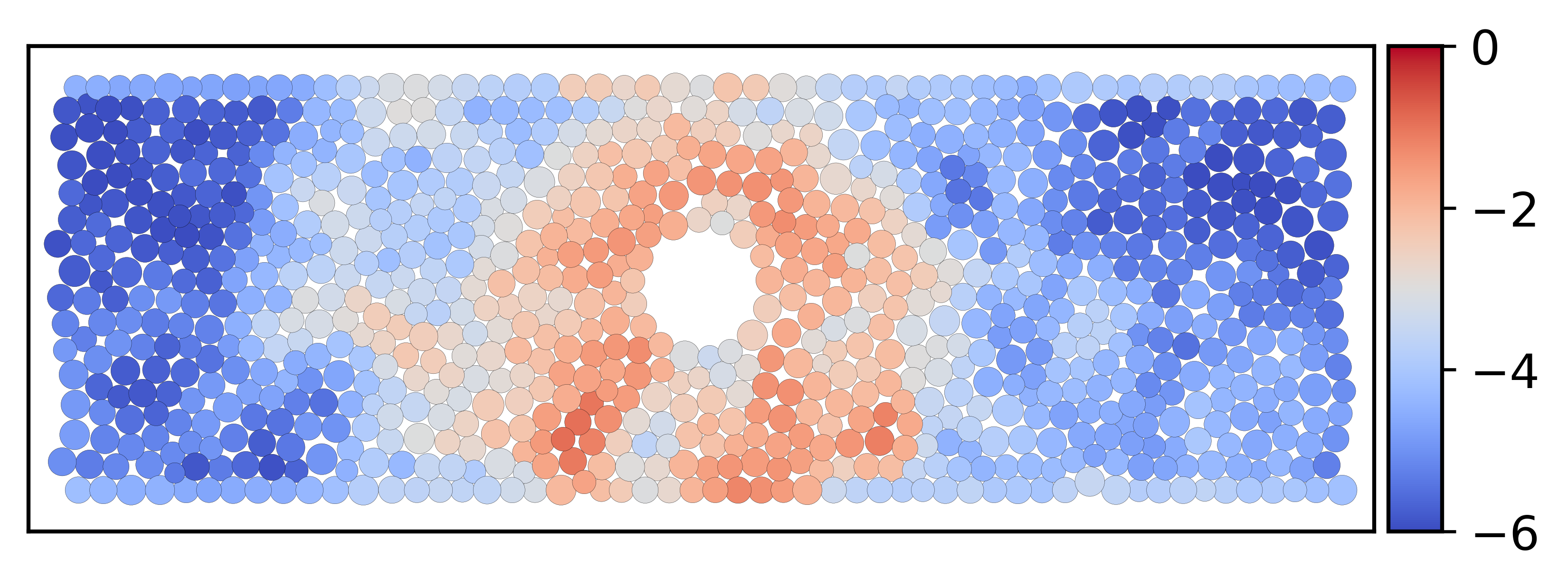}
        \label{fig:a}
    \end{subfigure}
    \hfill
    \begin{subfigure}{0.32\textwidth}
        \captionsetup{justification=Justified, singlelinecheck=false, position=above}
        \caption{Averaged $\log D_{\rm min}^2$ with $d=0.8$}
        \includegraphics[width=\textwidth]{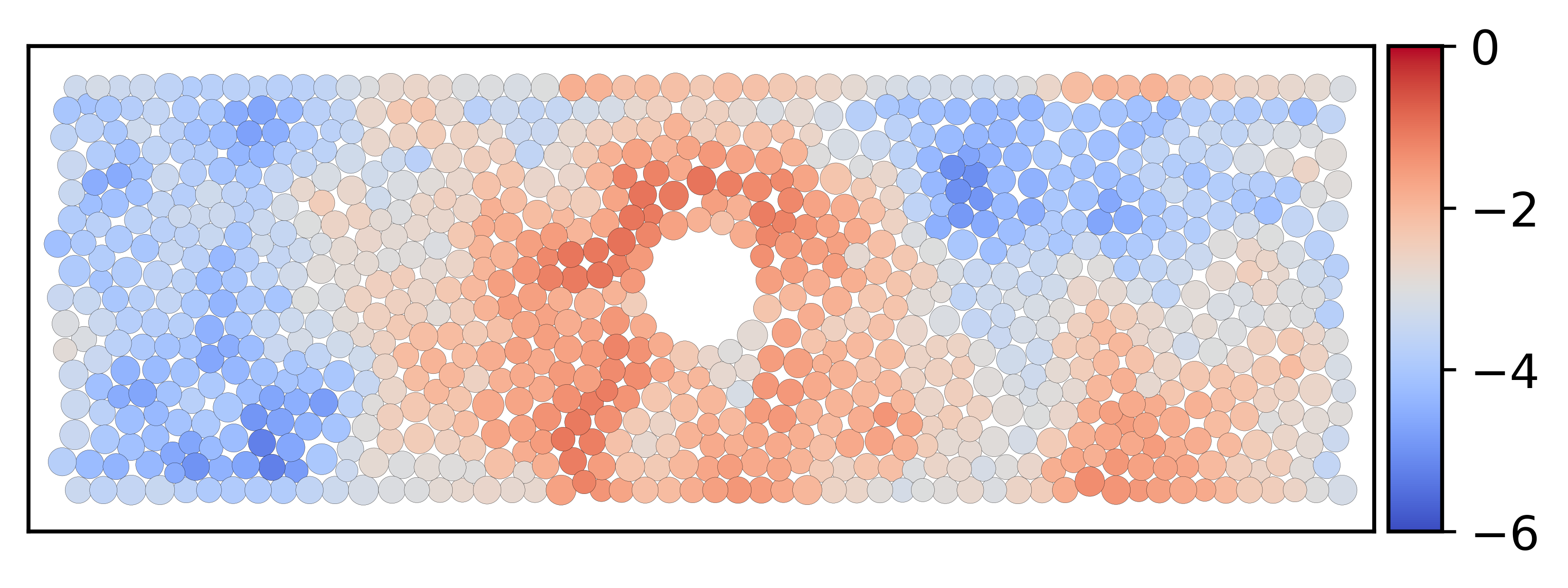}
        \label{fig:b}
    \end{subfigure}
    \hfill
    \begin{subfigure}{0.32\textwidth}
        \captionsetup{justification=Justified, singlelinecheck=false, position=above}
        \caption{Averaged $\log D_{\rm min}^2$ with $d=0.9$}
        \includegraphics[width=\textwidth]{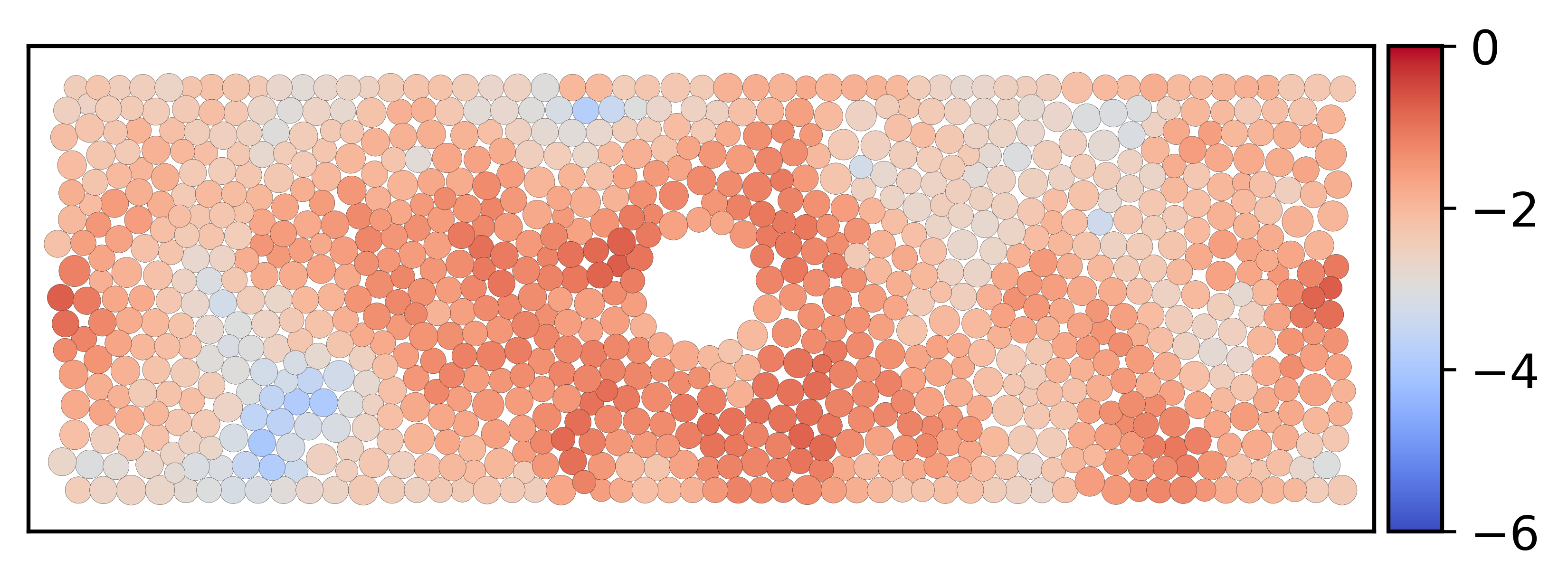}
        \label{fig:c}
    \end{subfigure}
    \hfill
    \begin{subfigure}{0.32\textwidth}
        \captionsetup{justification=Justified, singlelinecheck=false, position=above}
        \caption{Averaged $\log D_{\rm min}^2$ with $d=1.0$}
        \includegraphics[width=\textwidth]{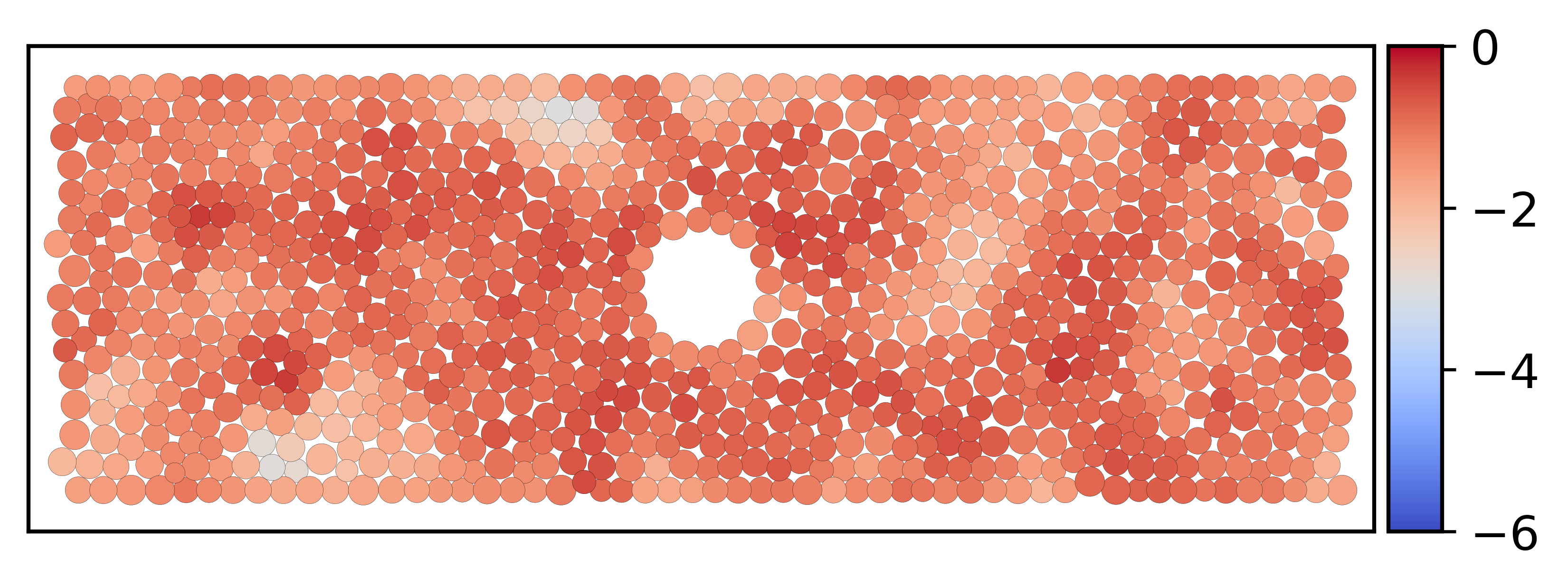}
        \label{fig:c}
    \end{subfigure}
    \caption{Visualization of the perturbation analysis for $\Delta t=300$. 
(a) Single-trajectory data of $\log D_{\rm min}^2$. 
(b--f) Mean $\log D_{\rm min}^2$ averaged over 10 trajectories perturbed with magnitude $d$.}
    \label{fig:perturbation}
\end{figure*}

In Fig.~\ref{fig:perturbation}, we compare the single-trajectory data of $\log D_{\rm min}^2$ with the mean $\log D_{\rm min}^2$ averaged over 10 perturbed trajectories. We find that, up to perturbation magnitudes $d\lesssim 0.5$, different trajectories generated from randomly perturbed initial configurations exhibit essentially the same heterogeneous pattern of plastic activity. Consequently, the trajectory-averaged field preserves the same heterogeneous pattern as the single-trajectory data. For larger values of $d$, the heterogeneous pattern is gradually destroyed, as expected.

Recall that lengths are measured in units of the mean particle diameter, $\overline{\sigma}=1$. Therefore, the fact that the same heterogeneous pattern is reproduced even for perturbations as large as $d=0.5$ suggests that the observed plastic heterogeneity is highly reproducible and not simply a chaotic consequence of tiny differences in the initial condition. This result indicates that the initial structure encodes information about the future heterogeneous dynamics.

This analysis implies that the limited ability of the present machine-learning model to predict the detailed heterogeneous pattern is mainly due to an insufficient characterization of the relevant local structural environment, rather than to an intrinsic unpredictability of the dynamics. More expressive approaches, including deep-learning-based methods~\cite{bapst2020unveiling,shiba2023botan,pezzicoli2024rotation} for extracting structural features, therefore represent a promising direction for future work.

\section{Conclusion and discussion}
\label{sec:conclusion}

In this work, we studied the prediction of plastic activity in the confined channel flow of amorphous soft particles around an obstacle using a two-dimensional computational bubble model. We considered two supervised-learning formulations: a regression task for the non-affine displacement $D^2_{\min}$ and a binary classification task for neighbor change events, which include conventional T1 rearrangements as a special case.

A key technical challenge arises from the presence of the obstacle and the confining walls, which explicitly break translational and rotational symmetries. To account for this symmetry breaking, we introduced additional input features encoding the positions of particles relative to the obstacle and the walls.

We systematically increased the complexity of the machine-learning models, starting from a simple linear baseline. We then examined the effects of a logarithmic transformation of the target variable, particle-size-dependent three-component descriptors, obstacle and wall descriptors, and local spatial averaging through coarse-graining. Finally, we employed a neural network to account for possible nonlinear relationships between the structural descriptors and future plastic activity.

We found that the obstacle and wall descriptors provide the largest improvement in predictive performance. Nevertheless, the machine-learning models considered here are still unable to reproduce the detailed heterogeneous pattern of plastic activity in individual configurations. We observed essentially the same limitation in both the regression and classification settings.

These results suggest two possible interpretations. First, the initial static structure may not contain sufficient information to determine the subsequent dynamics. Second, the machine-learning models and structural descriptors considered here may not be expressive enough to identify the structural features controlling the detailed spatial heterogeneity of plastic activity.

To distinguish between these possibilities, we performed an additional numerical analysis in which the initial particle configurations were perturbed and the subsequent dynamics were compared. We found that the heterogeneous pattern of future plastic activity remains remarkably robust against perturbations of the initial structure. This observation suggests that the future dynamics is, at least to some extent, encoded in the initial configuration and should therefore be predictable if the relevant structural information can be characterized sufficiently well. A natural next step is thus to employ more expressive deep-learning approaches, which have achieved substantial success in predicting the dynamics of glass-forming liquids~\cite{bapst2020unveiling,shiba2023botan,pezzicoli2024rotation}.

In this paper, we formulated the same prediction problem as both a regression task and a binary classification task. An interesting alternative would be to formulate it as an image-segmentation problem~\cite{long2015fully,ronneberger2015u}. In semantic image segmentation, a model assigns a class label to each spatial location in an image. Here, the spatial map of neighbor change events (e.g., Fig.~\ref{fig:dataset}(g,i)) could be represented as a binary field, with each pixel or particle-associated spatial region labeled according to whether a plastic rearrangement occurs. Segmentation architectures could then be used to predict the entire spatial pattern of plastic activity rather than treating each particle as an independent data point. 
%Because the present data are defined on irregular particle configurations, this approach would require either mapping the particle data onto a regular spatial grid or using segmentation methods designed for point clouds or graphs.

Finally, once sufficiently accurate predictions have been achieved for the computational model, an important direction will be to extend the approach to experimental datasets~\cite{dollet2007two}. Experimental data are generally more limited than simulation data. One possible strategy would therefore be to train the machine-learning model primarily on numerical simulations used as a digital twin of the experimental system and then transfer the learned representation to experimental data through transfer learning.

\begin{acknowledgments}
We thank Kirsten Martens for discussions.
We acknowledge support from MIAI@Grenoble Alpes and the Agence Nationale de la Recherche under the France 2030 (with the reference ANR-23-IACL-0006). 
This work was also supported by LabEx TEC21/UGA through the French National Research Agency in the framework of the ``France 2030'' program (ANR-15-IDEX-02).
\end{acknowledgments}

%\section*{Data Availability}The source code and dataset used in this paper are openly available at \href{https://github.com/alexandrestepanetz01-tech/Foam-dynamics-prediction} {https://github.com/alexandrestepanetz01-tech/Foam-dynamics-prediction}.

\appendix

\section{Behler-Parrinello (BP) structure descriptors}
\label{sec:BP}

We employ the Behler--Parrinello (BP) structural descriptors~\cite{behler2007generalized}, which have been widely used in various problems, including the prediction of glassy dynamics in two dimensions~\cite{cubuk2015identifying,rocks2021learning}. The BP descriptors consist of radial descriptors $G$ and angular descriptors $\Psi$.

\subsection{One-component descriptors}

We first ignore particle-size information and treat all particles as belonging to a single species. The radial descriptor $G_i$ for particle $i$ is defined as
\begin{equation}
    G_i = {\sum_j}' 
    \exp\left[-\frac{(r_{ij}-\mu)^2}{\ell^2}\right]
    f_c(r_{ij}),
\end{equation}
where $r_{ij}=|{\bf r}_i-{\bf r}_j|$ is the distance between particles $i$ and $j$, and $\mu$ and $\ell$ are parameters. The prime on the summation indicates that particle $i$ is excluded. The cutoff function is defined as
\begin{equation}
    f_c(r)=
    \begin{cases}
    \dfrac{1}{2}\left[\cos(\pi r/R_c)+1\right], & r \leq R_c,\\
    0, & r>R_c,
    \end{cases}
\end{equation}
where $R_c$ is the cutoff radius, which we set to $R_c=5.0$. We vary $\mu$ from $0.3$ to $5.0$ in increments of $0.1$, while fixing $\ell=0.1$, following Ref.~\cite{cubuk2015identifying}. This gives 48 radial descriptors. In this one-component description, the summation runs over all particles irrespective of their species.

The angular descriptor $\Psi_i$ for particle $i$ is defined as
\begin{eqnarray}
\Psi_i
&=&
2^{1-\zeta}
{\sum_{\substack{j,k\\ j\neq k}}}'
\exp\left[
-\frac{r_{ij}^2+r_{ik}^2+r_{jk}^2}{\xi^2}
\right] \nonumber \\
&\qquad& \qquad \times \left(1+\lambda\cos\theta_{ijk}\right)^\zeta
f_c(r_{ij})f_c(r_{ik})f_c(r_{jk}), \nonumber \\
\end{eqnarray}
where $\theta_{ijk}$ is the angle at particle $i$ formed by particles $i$, $j$, and $k$. The parameters $\xi$, $\lambda$, and $\zeta$ are varied systematically. We use the 22 parameter sets reported in Ref.~\cite{cubuk2015identifying}. Thus, the one-component BP representation contains $70$ descriptors in total: 48 radial descriptors and 22 angular descriptors.

\subsection{Three-component descriptors}

We next include particle-size information by distinguishing three species: small, medium, and large particles. We denote the corresponding sets of particles by S, M, and L, respectively. We also introduce species labels $A,B \in \{{\rm S},{\rm M},{\rm L}\}$ for the neighboring particles $j$ and $k$.

The species-resolved radial descriptor $G_i^A$ is defined as
\begin{equation}
    G_i^A =
    {\sum_{j\in A}}'
    \exp\left[-\frac{(r_{ij}-\mu)^2}{\ell^2}\right]
    f_c(r_{ij}) .
\end{equation}
The definitions of $r_{ij}$, $\mu$, $\ell$, and $f_c$ are the same as above. We use the same values of $\mu$, $\ell$, and $R_c$ as in the one-component case. Since there are three species, this gives $3\times48=144$ radial descriptors.

The species-resolved angular descriptor $\Psi_i^{AB}$ is defined as
\begin{eqnarray}
\Psi_i^{AB}
&=&
2^{1-\zeta}
{\sum_{\substack{j\in A,\ k\in B\\ j\neq k}}}'
\exp\left[
-\frac{r_{ij}^2+r_{ik}^2+r_{jk}^2}{\xi^2}
\right] \nonumber \\
&\qquad& \qquad \times
\left(1+\lambda\cos\theta_{ijk}\right)^\zeta
f_c(r_{ij})f_c(r_{ik})f_c(r_{jk}) . \nonumber \\
\end{eqnarray}
We use the same 22 parameter sets $(\xi,\lambda,\zeta)$ as in the one-component case. Since the unordered species pairs are SS, SM,\ SL, MM, ML, LL,
we obtain $6\times22=132$ angular descriptors. Thus, the three-component BP representation contains $144+132=276$ descriptors in total.

\subsection{Coarse-graining}

We consider coarse-graining of the three-component BP descriptors~\cite{jung2023predicting}. 
For $x_i=G_i^A$ or $\Psi_i^{AB}$, we define the coarse-grained descriptor as
\begin{equation}
    \overline{x}_i(\ell)
    =
    \frac{1}{\overline{\rho}_i(\ell)}
    \sum_{j\in\mathcal{N}_i}
    x_j
    \exp\left(-\frac{r_{ij}}{\ell}\right),
\end{equation}
where the local normalization factor is given by
\begin{equation}
    \overline{\rho}_i(\ell)
    =
    \sum_{j\in\mathcal{N}_i}
    \exp\left(-\frac{r_{ij}}{\ell}\right).
\end{equation}
Here, $\mathcal{N}_i$ denotes the set of particles included in the coarse-graining around particle $i$, including particle $i$ itself, and $\ell$ is the coarse-graining length scale. We set $\ell=2$.

\section{Training curves}
\label{sec:training_curves}

We monitor the training curves to assess whether overfitting occurs. 
This analysis is performed for linear regression, neural network, and logistic regression, using the corresponding performance score for each model as a function of the number of configurations in the training dataset.

\begin{figure*}[htbp]
    \centering
    \begin{subfigure}{0.32\textwidth}
        \captionsetup{justification=Justified, singlelinecheck=false, position=above}
        \caption{Linear regression}
        \includegraphics[width=\textwidth]{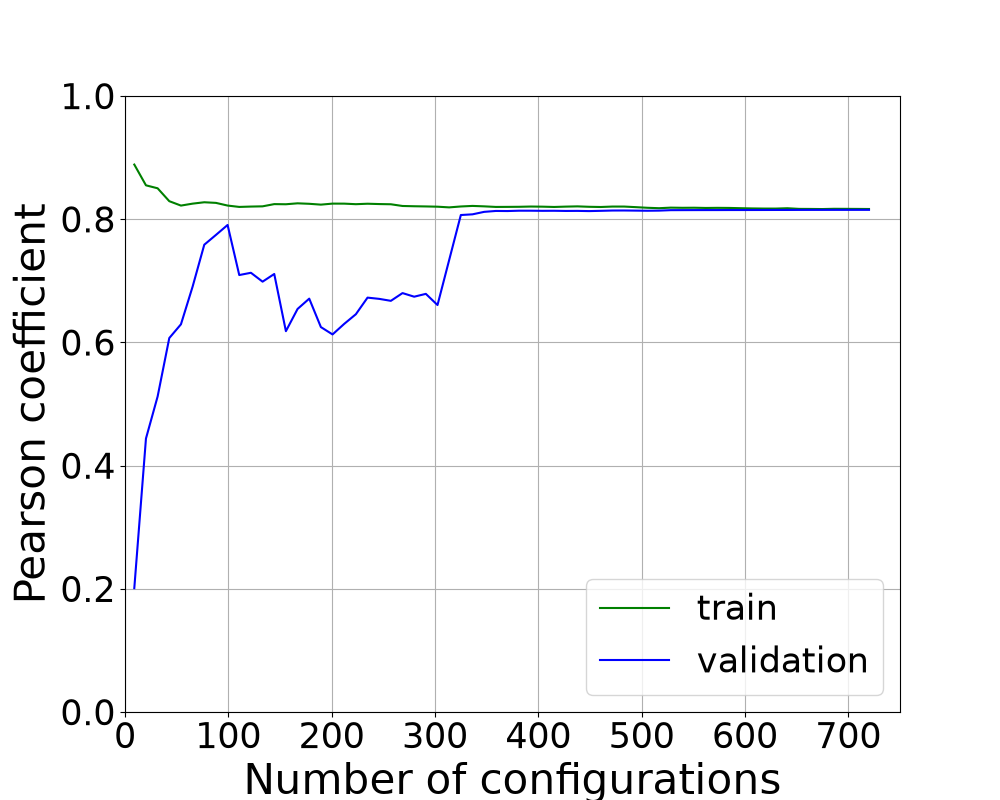}
        \label{fig:a}
    \end{subfigure}
    \hfill
    \begin{subfigure}{0.32\textwidth}
        \captionsetup{justification=Justified, singlelinecheck=false, position=above}
        \caption{Neural network}
        \includegraphics[width=\textwidth]{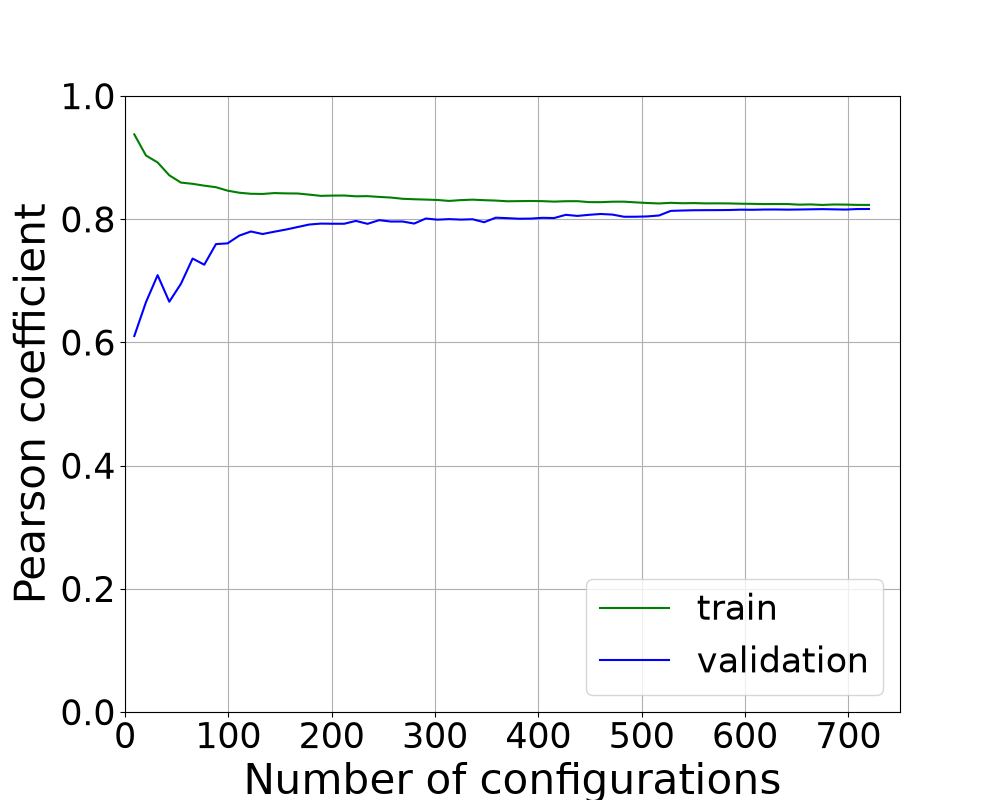}
        \label{fig:c}
    \end{subfigure}
    \hfill
    \begin{subfigure}{0.32\textwidth}
        \captionsetup{justification=Justified, singlelinecheck=false, position=above}
        \caption{Logistic regression}
        \includegraphics[width=\textwidth]{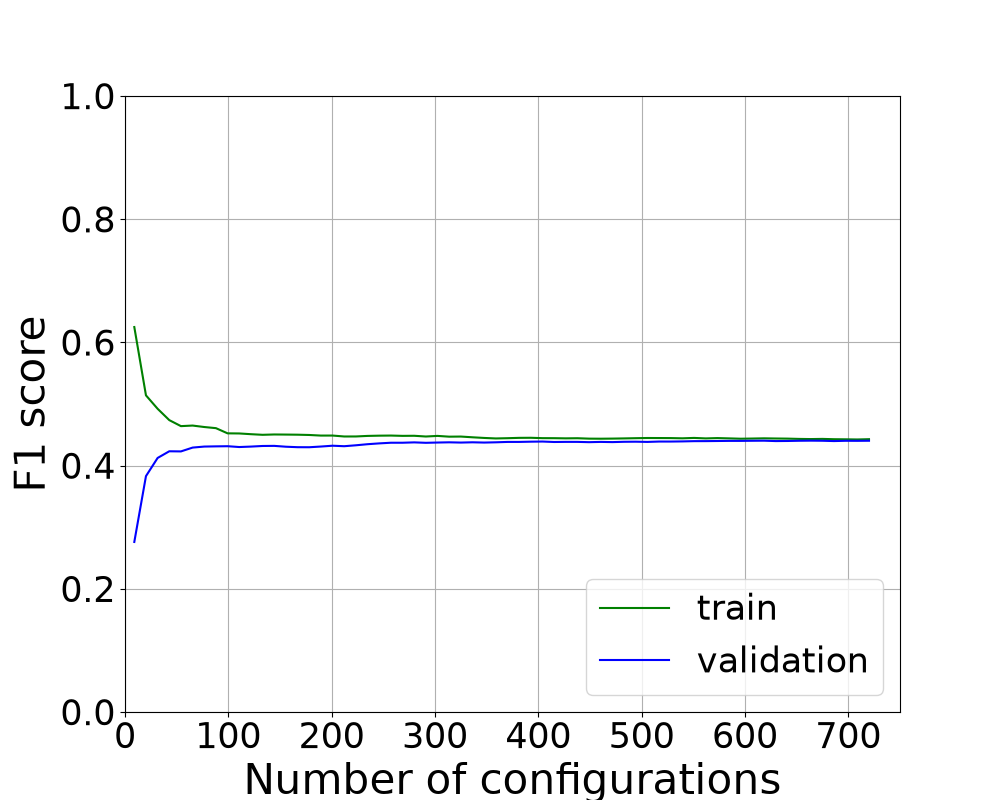}
        \label{fig:b}
    \end{subfigure}
    \caption{Training curves as a function of the number of training configurations for linear regression (a), neural network (b), and logistic regression (c) at the prediction timescale $\Delta t = 600$. 
The performance scores for the training and validation datasets are compared. For the neural network, the results shown are for the $(8,8)$ neural-network architecture.}
    \label{fig:training_curves}
\end{figure*}

The full dataset for one timescale $\Delta t$ contains 1000 configurations, which are split into 720 training configurations, 80 validation configurations, and 200 test configurations. 
We perform $K$-fold cross-validation and compare the performance on the training and validation datasets with $K=10$. 
The resulting training curves are shown in Fig.~\ref{fig:training_curves}. For all machine-learning models considered here, the training and validation performance scores converge to similar values as the number of training configurations increases. 
This behavior suggests that overfitting is negligible in our analysis.

\section{Classification performance}
\label{sec:class_performance}

We introduce the performance metrics used to evaluate neighbor change event prediction as a binary classification task~\cite{sokolova2009systematic}.
In the main text, we defined the neighbor change indicator, which takes the value 1 for a particle that undergoes a plastic rearrangement and 0 for a particle that does not.
Here, we refer to these two cases as the positive class, corresponding to label 1, and the negative class, corresponding to label 0.
The prediction results can be summarized by a $2\times 2$ confusion matrix. Its four
entries are defined as follows. A true positive ($TP$) is a particle that is predicted
to be positive and is indeed positive. A true negative ($TN$) is a particle that is
predicted to be negative and is indeed negative. A false positive ($FP$) is a particle
that is predicted to be positive but is actually negative. A false negative ($FN$) is a
particle that is predicted to be negative but is actually positive.

\begin{figure}
\includegraphics[width=0.9\linewidth]{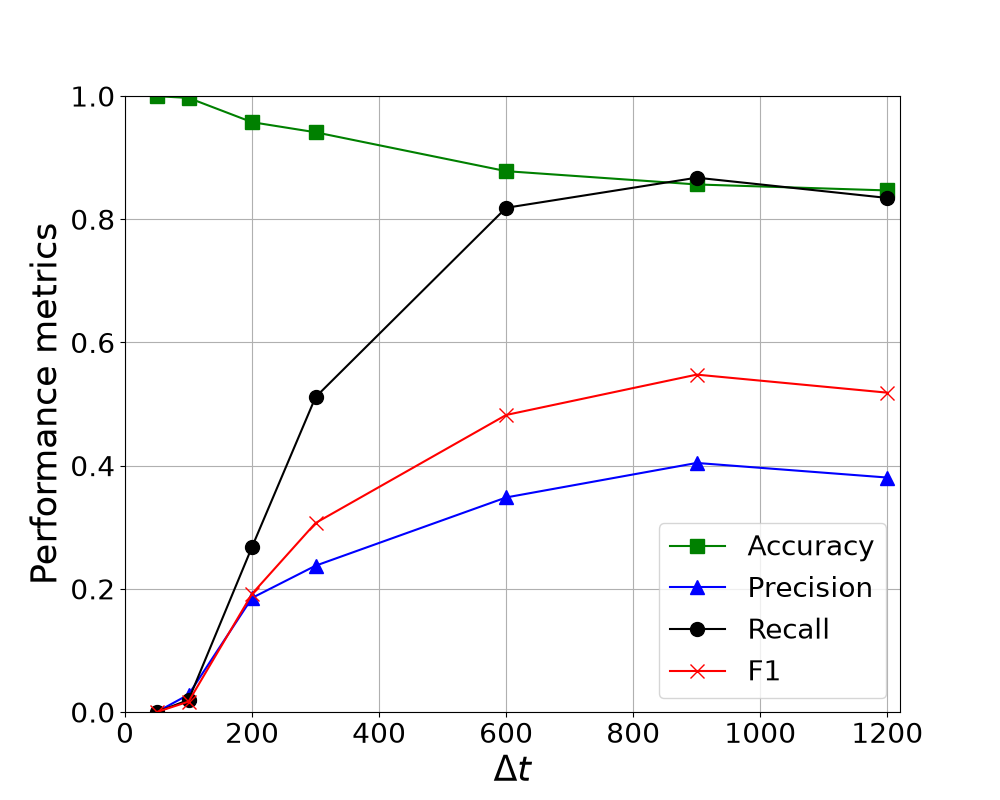}
\caption{Comparison of the four performance metrics used to evaluate the prediction of neighbor change events as a function of the prediction timescale $\Delta t$, using the same setup as the ``Coarse-grained descriptor'' case in Fig.~\ref{fig:performance_T1}.}
\label{fig:performance_metrics_T1_comparison}
\end{figure}

A simple performance metric is the accuracy,
\begin{equation}
    {\rm Accuracy}
    =
    \frac{TP+TN}{TP+FP+TN+FN}.
\end{equation}
Accuracy measures the fraction of correctly classified particles. However, it can be
misleading when the dataset is imbalanced. For example, if plastic events are rare,
a trivial model that predicts all particles as negative can still achieve high accuracy,
although it completely fails to detect plastic rearrangements.
As shown in Fig.~\ref{fig:performance_metrics_T1_comparison}, the accuracy score remains high for all $\Delta t$ because most particles belong to the negative class.

For this reason, we also use precision and recall. Precision is defined as
\begin{equation}
    {\rm Precision}
    =
    \frac{TP}{TP+FP}.
\end{equation}
It measures the reliability of positive predictions: among all particles predicted to
undergo a plastic rearrangement, precision tells us the fraction that actually do so.
A high precision means that the model makes few false alarms.

Recall is defined as
\begin{equation}
    {\rm Recall}
    =
    \frac{TP}{TP+FN}.
\end{equation}
It measures the ability of the model to detect actual positive events: among all
particles that truly undergo a plastic rearrangement, recall tells us the fraction
that are successfully detected. A high recall means that the model misses few plastic
events. When the denominator of precision or recall vanishes, the corresponding metric is set to zero, consistently with the \texttt{zero\_division=0} convention in scikit-learn.

In Fig.~\ref{fig:performance_metrics_T1_comparison}, we show both precision
and recall as functions of the prediction timescale $\Delta t$. Both quantities
increase at short times, reach a maximum at intermediate times, and then
decrease at longer times. Recall is systematically larger than precision,
whereas precision remains relatively small. This indicates that the model
detects a substantial fraction of actual neighbor change events, but also
produces a significant number of false positives.

Precision and recall are generally in a trade-off relation. If the classification
threshold is chosen so that the model predicts many particles as positive, recall
increases because more true plastic events are detected, but precision may decrease
because more false positives are also included. Conversely, if the threshold is chosen
more strictly, precision increases, but recall may decrease because some true plastic
events are missed.

To combine precision and recall into a single number, we use the F1 score,
\begin{equation}
    {\rm F1\ score}
    =
    \frac{2}{1/{\rm Precision}+1/{\rm Recall}}
    =
    \frac{2TP}{2TP+FP+FN}.
\end{equation}
The F1 score is the harmonic mean of precision and recall. It becomes large only when
both precision and recall are large, and is therefore a useful metric for evaluating
the prediction of rare plastic events.

Figure~\ref{fig:performance_metrics_T1_comparison} also shows the F1 score.
During the time evolution, the F1 score takes values between precision and
recall.

\bibliography{refs_foam_projects}% Produces the bibliography via BibTeX.

\end{document}